\documentclass[showpacs,showkeys,aps,prd,nofootinbib,floatfix,amsmath,amssymb]{revtex4}
\usepackage{amssymb}
\usepackage{amsmath}
\usepackage{amsfonts}
\usepackage{epigraph}
\usepackage{mathrsfs}
\usepackage{graphics}
\usepackage{bm}
\usepackage{bbm}	% Math Blackboard
\usepackage{tensor} 	% Tensor indices
\usepackage{slashed}	% Dirac Slash Notation 
\usepackage{tikz,ifthen}% TikZ!
\usepackage[toc,page]{appendix}

\usepackage{tikz}
\usetikzlibrary{trees}
\usetikzlibrary{decorations.pathmorphing}
\usetikzlibrary{decorations.markings}
\usepackage{tikz,ifthen}% TikZ!
\tikzset{beamerprimary/.style={structure.fg, thick}}
\tikzset{beamersecondary/.style={structure.bg, thick}}
\tikzset{boson/.style={draw=structure.fg,decorate, decoration={snake}},
    gauge/.style={decorate, decoration={snake} },
    fermion/.style={postaction={decorate},
        decoration={markings,mark=at position .55 with {\arrow{>}}}},
    fermionloop/.style={postaction={decorate},
        decoration={markings,mark=at position .25 with {\arrow{<}}}}, 
    gluon/.style={decorate, 
        decoration={coil,amplitude=4pt, segment length=5pt}},
    scalar/.style={dashed},
    graviton/.style={double}	
}

\newcommand{\bpsi}{\bar{\psi}}

\begin{document}

\title{Renormalization of the QED of second order spin $\frac{1}{2}$ fermions.}
\author{Ren\'e \'Angeles-Mart\'{i}nez, Mauro Napsuciale }
\affiliation{Departamento de F\'{i}sica, Universidad de Guanajuato, Lomas del Campestre 103, Fraccionamiento 
Lomas del Campestre, Le\'on, Guanajuato M\'exico, 37150.}

\begin{abstract}
In this work we study the renormalization of the electrodynamics of spin $1/2$ fermions in the Poincar\'e 
projector formalism which is second order in the derivatives of the fields. We analyze the superficial degree of 
divergence of the vertex functions of this theory, calculate at one-loop level the vacuum polarization, 
fermion self-energy and $\gamma-fermion-fermion$ vertex function and the divergent piece of the one-loop 
contributions to the $\gamma-\gamma-fermion-fermion$  vertex function. It is shown that 
these functions are renormalizable independently of the value of the gyromagnetic factor $g$ which is a free 
parameter of the theory. We find a photon propagator and a running coupling constant $\alpha (q^2)$ that depend 
on the value of $g$. The magnetic moment form factor contains a divergence associated to $g$ which disappears 
for $g=2$ but in general requires the coupling $g$ to be renormalized. A suitable choice of the renormalization 
condition for the magnetic form factor yields the one loop finite correction $\Delta{g}=g\alpha/2\pi$. 
For a particle with $g=2$ we recover results of Dirac theory for the photon propagator, the running 
of $\alpha (q^2)$ and the one-loop corrections to the gyromagnetic factor. 
\end{abstract}
\keywords{Renormalization, electromagnetic properties.}
\pacs{11.10.Gh,11.30.Cp,13.40.Em}
\maketitle

%\tableofcontents

\section{Introduction}

The proper description of interacting high spin fields has been addressed by many authors and we are 
still awaiting for conclusive results. In fact, after the formulation of the Rarita-Schwinger 
formalism, it was clear that the corresponding interacting high spin fields suffer from serious inconsistencies  
\cite{superluminal}. Recently, a possible solution was suggested based on the projection onto eigensubspaces 
of the Casimir operators of the Poincar\'e group \cite{NKR}.
Indeed, in \cite{NKR} the case of the propagation of spin 3/2 interacting fields 
in the $(1/2,1/2)\otimes [(1/2,0)\oplus (0,1/2)]$ was addressed 
in detail and it was shown there that there is a deep connection between the causal  
propagation of spin 3/2 waves and the specific value $g=2$ for the gyromagnetic factor of the spin 
3/2 particle. Later on, it was shown that the same value is related to the unitarity of the 
Compton scattering amplitude in the forward direction \cite{DN}.

In order to gain insight into the formal structure of the formalism, the case of spin 1 in the (1/2,1/2) 
representation was studied in \cite{NRDK}. In this case the most general electromagnetic interaction of the 
spin 1 vector particle was also shown to depend on two parameters, the gyromagnetic factor $g$ and a 
parameter denoted by $\xi$ associated to parity violating interactions, which 
cannot be fixed from the Poincar\'{e} projection alone. These parameters determine the electromagnetic 
structure of the particle and were fixed imposing unitarity at high energies for Compton scattering. 
This procedure fixes the parameters to $g=2$ and $\xi=0$ predicting a gyromagnetic factor 
$g=2$, a related quadrupole electric moment $Q=-e(g-1)/m^{2}$ and vanishing odd-parity
couplings as a consequence of $\xi=0$  . The obtained couplings coincide with the ones 
predicted for the  $W$ boson in the Standard Model. 

The simplest spin 1/2 case in the $(1/2,0)\oplus(0,1/2)$ representation was addressed in \cite{DNR}. 
This case is interesting at least in the formulation of effective field theories for the electromagnetic 
properties of hadrons where the low energy constants are precisely the free parameters in the Lagrangian.
Indeed, the electromagnetic interactions of a spin 1/2 fermion also depend on two free parameters, 
the gyromagnetic factor $g$ and a parameter $\xi$ related to odd-parity Lorentz structures.  
A calculation of Compton scattering in this formalism yields similar results to Dirac theory in the 
particular case $g=2,~\xi=0$ and for states with well defined parity.  

In all the studied cases of spin $1/2,1,3/2$, we find the correct classical limit and a finite value $r_{c}^{2}=\alpha/m$ 
for the differential cross section in the forward direction, independently of the photon energy and of the value of the free 
parameters, the same value as in scalar electrodynamics.

These results motivate us to study the renormalization of the Poincar\'e projector formalism. In order 
to understand possible difficulties of the quantum theory we start here with the technically simplest case of spin 1/2. 

A second order formalism for the description of spin $1/2$ fermions was considered by Feynman in an appendix 
of \cite{feynman1}, following a seminal work by V. Fock \cite{Fock}. Some years later, the V-A structure 
of the weak interactions was motivated by Feynman and Gell-Mann based on the equation of motion obtained 
by decomposing the Dirac wave function interacting with an electromagnetic background into its Weyl components \cite{fg}. 
The resulting equation for the interacting Weyl wave function turns out to be of second order in the derivatives 
of the two-component spinors. An additional motivation to follow this idea was the simplicity 
of the evaluation of the corresponding path integrals with second order fermions which 
is presently useful in the word line formulation of perturbative quantum field theory \cite{schubert}. 

After Feynman and Gell-Mann proposed their equation, the  relativistic quantum mechanics aspects were studied in  
\cite{hanfeygellclasico}, \cite{cufarofeygellclasico1}, \cite{cufarofeygellclasico2}, \cite{hosclasico1}, 
\cite{hosclasico2}. The corresponding quantum field theory was also considered and applied in 
the calculation of some processes \cite{laurie}, \cite{tonin}, \cite{hebert}, \cite{3order}, \cite{volkovyskii}, \cite{hos1}. 
The non-abelian version of this formalism was studied in \cite{hos2}. The possibility that second order fermions  
avoid the problems of chiral fermions on the lattice were studied in \cite{Longhitano}, \cite{laticepalumbohiggs4}. 
Recent discussions of the formalism for non-abelian and abelian fields can be found in  
\cite{morgan}, \cite{veltman}. At one-loop level  there are some partial results in \cite{laurie},
\cite{hebert}, \cite{hos1}, \cite{Longhitano}. Specially in \cite{hebert} the divergent part of the one-loop contributions 
to the two- and three-point functions are isolated, these vertex functions are proved to be renormalizable 
and the one-loop correction to the magnetic moment is shown to coincide with the result of the Dirac theory.

In contrast to the Feynman-Gell-Mann formalism which is a careful rewriting of the Dirac equation, the 
Poincar\'e projector formalism starts from a different but basic principle: the projection onto well defined subspaces 
of the Poincar\'e Casimir operators in a given Lorentz representation, which fixes only the Poincar\'e good 
quantum numbers, the mass and spin of the particle and yields a  more general structure allowing 
for arbitrary values of the gyromagnetic factor.   

In this work we study the one-loop level structure of the electrodynamics of spin $1/2$ fermions in  
the Poincar\'e projector formalism. We analyze the superficial degree of divergence of the vertex functions 
and perform a complete calculation of the 2- and 3- point functions at one loop level. We go a step forward and 
 calculate the divergent piece of the $\gamma-\gamma-fermion-fermion$ ($\gamma\gamma ff $)vertex function. 
It is shown that this  vertex function is renormalizable for arbitrary values of the gyromagnetic factor. 

This paper is organized as follows. In the next section we present the 
Feynman rules and the derivation of the Ward-Takahashi identities used in the paper. In section III  we carry 
out the renormalization procedure. We analyze the superficial degree of divergence of the vertex functions, 
rewrite the Lagrangian in terms of the renormalized parameters,  calculate the one-loop corrections to 
the propagators and the three-point vertex function and show that the  $\gamma\gamma ff$ 
vertex function is renormalizable at one-loop level. A summary of our results is given in section IV. 
Details of the Lorentz structure, its $d$-dimensional extension and of scalar functions arising in the calculation 
of the three-point vertex function are given in the appendix.

\section{Feynman rules and Ward-Takahasi identities}

The generating functional for the Green functions of the Poincar\'e
projector formalism for spin $1/2$ fermions is
\begin{equation}
Z[J,\eta,\bar{\eta}]= N\int \mathscr{D}A_\mu \mathscr{D}\bar{\psi}
\mathscr{D}\psi \text{exp}{\Big[ i\int \mathscr{L} dx \Big]},
\label{genfunc}
\end{equation}
with \cite{DNR}
\begin{equation}
\mathscr{L}=  -\frac{1}{4}F^{\mu\nu}F_{\mu\nu}- \frac{1}{2\alpha}(
\partial^\mu A_\mu)^2+ D^{\dagger\mu} \bar{\psi}
T_{\mu\nu}D^\nu\psi-m^2\bar{\psi}\psi +J^{\mu} A_{\mu}+
\bar{\eta}\psi+\bar{\psi}\eta .
\label{NKRlag}
\end{equation}
Here  $D_\mu= \partial_\mu+ie A_\mu $ (fermion charge $-e$) stands for the
covariant derivative, $\eta , \bar{\eta}$
are the fermionic external currents and the space-time tensor $T^{\mu\nu}$
is given by
\begin{equation}
T^{\mu\nu}\equiv g^{\mu\nu}- (ig -  \xi \gamma^5)M^{\mu\nu},
\end{equation}
where $M_{\mu \nu}$ stands for the generators of the $(\frac{1}{2},0)\oplus (0,\frac{1}{2})$ representation 
of the Lorentz group. The free parameters of the theory, besides $e$ and $m$, are the gyromagnetic factor $g$ and 
the parameter $\xi$ related to parity violating interactions . A straightforward calculation yields the 
Feynman rules in Fig. \ref{FD}, where we use the Feynman gauge.
\begin{figure}[ht]
\begin{tikzpicture}
%propagators
\draw[fermion] (-6,4)  -- node[above]{$p$}(-4,4) ;
\node at (-5,3) {$iS(p) =\frac{i}{p^{2}-m^{2}+i\varepsilon} \equiv\frac{i}{\square \left[ p\right] }$};
\draw[decorate,decoration=snake] (0,4) -- node[above]{$q$}(2,4);
\node at (1,3) {$ i\Delta_{\mu\nu}(q) =
\frac{-ig_{\mu \nu }}{q^{2}+i\varepsilon } \equiv \frac{-ig_{\mu \nu }}{\triangle
(q)}   $};
\node at (0,4.2) {$\mu $};
\node at (2,4.2) {$\nu $};
%vertex
\draw[fermion] (-6,-1) node[above]{$p$} --(-5,0) ;
\draw[fermion] (-5,0)  --  (-4,-1) node[above]{$p^{\prime}$};
\draw[decorate,decoration=snake] (-5,1)--(-5,0);
\node at (-5.5,1.0) {$q, \mu $};
\node at (-6,-2.0) {$-ieV_{\mu }(p,p^{\prime}) =-ie
\left[(p^{\prime}+p)_{\mu }+(ig+\xi\gamma^5) M_{\mu \nu }(p^{\prime
}-p)^{\nu}\right] $};
%seagull
\draw[decorate,decoration=snake] (0,1)--(1,0);
\draw[decorate,decoration=snake] (1,0)--(2,1);
\draw[fermion] (0,-1) node[above]{$p$} --(1,0) ;
\draw[fermion] (1,0)  --  (2,-1) node[above]{$p^{\prime}$};
\node at (0.3,1) {$\mu $};
\node at (1.8,1) {$\nu $};
\node at (1,-2) { $\,\,\,\,\,\,\, ie^2V_{\mu \nu}(p,q,p',q')=2ie^{2} g_{\mu \nu } $};
\end{tikzpicture}
\caption{Feynman rules for the QED of second order fermions in the Feynman
gauge $\alpha=1$.}
\label{FD}
\end{figure}

Gauge invariance imposes relations among different Green functions. These relations will be used
below as cross checks of our calculations, thus we sketch their derivation here.
Under an infinitesimal gauge transformation $A_\mu \to A_\mu +\partial_\mu \Lambda $,
$\psi\to \psi-ie\Lambda\psi$, $\bar{\psi}\to
\bar{\psi}+ie\Lambda\bar{\psi}$, the Lagrangian transforms as
\begin{equation}
\mathscr{L} \to \mathscr{L}-   (\partial _\mu A^\mu) \square \Lambda
+J^\mu \partial_\mu \Lambda-ie\Lambda \bar{\eta}\psi +ie\Lambda \bpsi \eta.
\end{equation}
The variation of the generating functional must vanish which implies that
\begin{align}
\Big[    i \frac{\partial^{2}}{\alpha} (\partial^\mu \frac{\delta}{\delta
J^\mu(x)} )
-\partial_\mu J^\mu -e(\bar{\eta}\frac{\delta}{\delta\bar{\eta}(x)} + \eta
\frac{\delta}{\delta\eta(x)} ) \Big]
Z[J,\eta,\bar{\eta}] =0.
\label{ZLambda}
\end{align}
In terms of the generating functional for connected diagrams $W[J,\eta,\bar{\eta}]$ 
which is related to $Z[J,\eta,\bar{\eta}]$ by
\begin{align}
Z[J,\eta,\bar{\eta}]= e^{iW [J,\eta,\bar{\eta}]}= \sum_N \frac{i^N}{N!}
[W[J,\eta,\bar{\eta}]]^N,
\end{align}
equation (\ref{ZLambda}) can be rewritten as
\begin{align}
-\frac{\partial^{2} }{\alpha} \partial^\mu \frac{\delta W}{\delta J^\mu} -
 \partial^\mu J_{\mu}
-ie \Big[ \bar{\eta} \frac{\delta W}{\delta  \bar{\eta}} + \eta 
\frac{\delta W}{\delta \eta}  \Big]=0.
\label{ward}
\end{align}
Writing now this equation in terms of the following function
\begin{equation}
i\Gamma[\psi,\bar{\psi},A_{\mu}]=iW[J,\eta,\bar{\eta}]- i\int dx
(\bar{\eta} \psi+\bar{\psi} \eta+J^{\mu}A_{\mu}),
\end{equation}
we get
\begin{align}
-\frac{\partial^{2}}{\alpha} \partial^{\mu} A_{\mu}(x) +\partial_{\mu} \frac{\delta
\Gamma}{\delta A_{\mu}(x)}
+ ie \frac{\delta \Gamma}{\delta  \psi(x)}\psi + ie \frac{\delta
\Gamma}{\delta \bar{\psi}(x)} \bar{\psi}= 0.
\label{WI}
\end{align}
This is the master relation for Ward Identities in configuration space.
Using successive functional derivatives
with respect to the fields at different space-time points and evaluating
at zero fields we get relations
among distinct vertex functions. As an example we take the functional
derivatives with respect to
$\bar{\psi}(x_1)$  and  $\psi(y_1)$ and evaluating at $A_{\mu}= 0,
~\psi=0, ~\bar{\psi}=0$ we get the first Ward-Takahashi identity in configuration space
\begin{equation}
\partial^\mu \frac{\delta^3 \Gamma[0]}{\delta \bar{\psi}(x_1) \delta
\psi(y_1) \delta A^{\mu}(x)}=
ie \delta(x-y_1)\frac{\delta^2 \Gamma[0]}{\delta \bar{\psi}(x_1) \delta
\psi(x) }
 - ie\delta(x- x_1) \frac{\delta^2 \Gamma[0]}{    \delta \bar{\psi}(x)
\delta \psi(y_1) } . \label{WI1}
 \end{equation}
This relation is more useful in momentum space. We denote by
$\Gamma^\mu(p,q,p')$ the $\gamma-fermion-fermion$ ($\gamma ff$) irreducible vertex
in momentum space and by $S'^{-1}(p)$ the inverse exact propagator in the
presence of interactions
 \begin{eqnarray}
\int dxdy_1 dx_1 e^{-i(xq+py_1-p'x_1)}  \frac{\delta^3 \Gamma[0]}{\delta
\bar{\psi}(x_1) \delta \psi(y_1) \delta A^{\mu}(x)}
&=&ie(2\pi)^4 \delta(p'-p-q)   \Gamma_\mu(p,q,p'), \\
\int dx_1 dy_1  e^{-i(py_1-p'x_1) } \frac{\Gamma[0]}{\delta
\bar{\psi}(x_1) \delta \psi(y_1) }
&=& (2\pi)^4 \delta(p'-p) S'^{-1}(p).\label{1iward}
\end{eqnarray}
Fourier transforming \eqref{WI1} we obtain the first Ward identity in
momentum space
\begin{align}
q^\mu \Gamma_{\mu}(p,q,p+q)= S'^{-1}(p+q)-S'^{-1}(p).
\label{1iw}
\end{align}
A differential form of this equation can be obtained taking $q\to 0$
\begin{align}
\Gamma_{\mu}(p,0,p)= \frac{\partial S'^{-1}(p)}{\partial p^\mu }.
\label{1iwd}
\end{align}
This identity  must be satisfied to any order in perturbation theory. From
the Feynman rules in
Fig. \ref{FD} we can easily check that it holds at tree level.

Similar calculations using the third order functional derivative
$\frac{\delta^3 }{\delta \bar{\psi}(x_1)\delta \psi(y_1) \delta A_\nu(y)}$
 on Eq. (\ref{WI}) allow us to
derive the following Ward-Takahashi identity relating the $\gamma\gamma ff$ to
the $\gamma ff$ vertex function as
\begin{align}
q^\mu \Gamma_{\mu\nu}(p,q,p',q')=
\Gamma_\nu(p+q,q',p')-\Gamma_\nu(p,q',p'-q)\label{2iward},
\end{align}
whose differential form is
\begin{align}\label{2iwd}
 \Gamma_{\mu\nu}(p,0,p',q')=\frac{\partial \Gamma_\nu(p,q',p')}{\partial
p^\mu} + \frac{\partial\Gamma_\nu(p,q',p')}{\partial p'^\mu}.
\end{align}
Again, the tree level vertices $\Gamma^{(0)}_{\mu\nu}(p,q,p',q')= V_{\mu\nu}(p,q,p',q')$ and
$\Gamma^{(0)}_{\mu}(p,q,p')=V_{\mu}(p,p')$ in Fig. (\ref{FD}) satisfy these relations.

\section{Renormalization} 

\subsection{Superficial degree of ultraviolet divergences}

In general, the calculation of a diagram connecting certain number of initial and final particles 
involves integrals with the following generic form
\begin{eqnarray}
I= \int d^4l_1 ...d^4l_n \frac{\tau_{\mu\nu...}(l_1,...,l_n,...)}{\triangle[l_i...] ...\square[l_j...]}.
\end{eqnarray}
The superficial degree of divergence  of these integrals is defined as 
\begin{equation}
D=N_{l}-D_{l}+4n_{l},
\end{equation}
where $N_{l}$ stands for the number of powers of loop momenta of the diagram in the numerator, $D_{l}$ denotes 
the number of powers of the loop momenta in the denominator and $n_{l}$ represents the number of independent 
loop momenta in the integral.  In the ultraviolet region all momenta are large enough to disregard the constants 
in the integral which behaves like 
\begin{align}
\int^\infty l^{D-1} dl.
\end{align}
If $D=0$ we say that the integral is logarithmically divergent. In the case $D=1$ we refer to it as linearly divergent and 
for negative $D$ the integral is convergent. A renormalizable theory requires a Lagrangian with dimensionless 
couplings and to have a limited number of divergent diagrams which can be re-absorbed in the definitions of the 
parameters (masses and couplings) of the theory. 

The action for the QED of second order fermions in four dimensions is
\begin{align}
I = \int d^4x \mathscr{L}= \int d^4x\big[ - \frac{1}{4}F^{\mu\nu}F_{\mu\nu} 
- \frac{1}{2\alpha}(\partial^\mu A_\mu)^2+ D_\mu \bpsi  
T^{\mu\nu} D_\nu \psi- m^2\bpsi \psi \big],
\end{align}
where $D_\mu= \partial_\mu +ie A^\mu$. Notice that a dimensionless action requires the fermion fields to have 
dimension $1$ in four dimensions ($\frac{d-2}{2}$, for dimension $d$), same dimension as the gauge fields. 

For an arbitrary connected Feynman diagram we use the following definitions: 
$L\equiv$ number of loops, $P_i \equiv $ number of photon internal lines, $E_i \equiv $ number of fermion internal lines, 
$P_e\equiv $ number of photon external lines, $E_e\equiv $ number of fermion external lines, $n_3\equiv$ number of   
$\gamma ff$ vertices and $n_4\equiv$ number of $\gamma\gamma ff$ vertices. In a given integral, all propagators contribute
with dimension $l^{-2}$ to the integral, $\gamma ff$ vertices contribute at  
most with a factor $l$ and the $\gamma\gamma ff$ vertex does not increase the degree of divergence which is given by 
\begin{eqnarray}
D\le 4 L - 2P_i- 2E_i +n_3 . \label{dsd1}
\end{eqnarray}
Furthermore, we have momentum conservation both global and for each vertex which requires 
\begin{eqnarray}
L= E_i+P_i-n_3-n_4+1 .  \label{dsd2}
\end{eqnarray}
In addition, the vertices $\gamma ff$ and $\gamma\gamma ff$ are connected to  two fermionic lines thus 
\begin{eqnarray}
2(n_3+n_4)= E_e+2E_i. \label{dsd3}
\end{eqnarray}
Finally the  $\gamma ff$ vertex connects always to a photonic line while the $\gamma\gamma ff$ vertex connects 
to two photonic lines  which imposes the following relation
\begin{eqnarray}
2n_4+n_3= P_e+2P_i. \label{dsd4}
\end{eqnarray}
Using Eq. (\ref{dsd2}) in Eq. (\ref{dsd1}) and replacing  $E_i, P_i$ as obtained from Eqs. (\ref{dsd3},\ref{dsd4}) we 
obtain  
\begin{eqnarray}
D\le 4- E_e-P_e. \label{divsup}
\end{eqnarray}
The superficial degree of divergence is then dictated only by the number of external lines. We get at most 
quadratic divergences for the two point functions, linear divergences for the three point functions and 
logarithmic divergences for the four point functions. All connected diagrams with more than four external lines are 
convergent. 

\subsection{Counterterms}
In this work we will carry out the renormalization procedure in the case of $\xi=0$, i.e. in the 
case of vanishing odd parity interactions. The calculation of quantum corrections to parity violating interactions 
requires to consider the problem of the proper definition of chirality in dimension $d$ which is beyond the scope 
of this work. In the case $\xi=0$ the parameters in the bare Lagrangian are the fermion mass $m_d$,  
the fermion charge $e_d$, and the gyromagnetic factor $g_d$.  The renormalized fields are related to the bare ones as 
\begin{equation}
A^{\mu}_r =  Z_1^{-\frac{1}{2}} A^{\mu}_d , \qquad  \psi_r =  Z_2^{-\frac{1}{2}} \psi_d.
\end{equation}
It is convenient to split the Lagrangian into its free and interacting parts  
 \begin{equation}
 \mathscr{L}=\mathscr{L}_0+\mathscr{L}_i ,
 \end{equation}
where 
\begin{align}
\mathscr{L}_0=-\frac{1}{4}F^{\mu\nu}_{d}F_{d\mu\nu}- \frac{1}{2}(\partial^\mu A_{d\mu})^2
+ \partial^\mu \bpsi_d \partial_{\mu} \psi_d -m_{d}^{2} \bpsi_d \psi_d,\\
\mathscr{L}_i= -ie_d [ \bpsi_d   T_{d\nu\mu} \partial^\mu \psi_d - \partial^\mu 
\bpsi_d T_{d\mu\nu}\psi_d   ] A^\nu_d + e^{2}_d \bpsi_d  \psi_d A^\mu_d A_{d\mu},
\end{align}
with 
\begin{equation}
T^{\mu\nu}_{d}\equiv  g^{\mu\nu}-ig_d M^{\mu\nu}.
\end{equation}
Writing the Lagrangian in terms of the renormalized fields we get the free Lagrangian as 
\begin{align}
\mathscr{L}_0=&-\frac{1}{4}F^{\mu\nu}_r F_{r\mu\nu}  - \frac{1}{2}(\partial^\mu A_{r\mu})^2
-\frac{1}{4}F^{\mu\nu}_r F_{r\mu\nu} \delta_{Z_1}  - \frac{1}{2}(\partial^\mu A_{r\mu})^2 \delta_{Z_1} \\%
&+ \partial^\mu \bpsi_r \partial_{\mu} \psi_r   - m_{r}^{2} \bpsi_r \psi_r 
+[ \partial^\mu \bpsi_r \partial_{\mu} \psi_r   - m^2   \bpsi_r \psi_r ] \delta_{Z_2} 
- \delta_{m}   \bpsi_r \psi_r ,
\end{align}
where we used the following definitions 
\begin{equation}
\delta_{Z_1}\equiv Z_{1}-1, \qquad \delta_{Z_2}\equiv Z_{2}-1,  \qquad \delta _{m}\equiv Z_{2}[ m^2_d -m^2_r].
\end{equation}
Similarly, the interacting Lagrangian can be rewritten as 
\begin{align*}
\mathscr{L}_i  = &-ie_r [ \bpsi_r   T_{r\nu\mu} \partial^\mu \psi_r - \partial^\mu \bpsi_r 
T_{r\mu\nu} \psi_r   ] A^\nu_r    -ie_r [ \bpsi_r   T_{r\nu\mu} \partial^\mu \psi_r - 
\partial^\mu \bpsi_r T_{r\mu\nu} \psi_r   ] A^\nu_r \delta_{e} \\ %
& -ie_r [ \bpsi_r   (-ig_r M_{\nu\mu})  \partial^\mu \psi_r - \partial^\mu \bpsi_r 
(-ig_rM_{\mu\nu})\psi_r   ] A^\nu_r   \delta_{g} + e^{2}_r \bpsi_r  \psi_r A^\mu_r A_{r\mu}
+  e^{2}_r \bpsi_r  \psi_r A^\mu_r A_{r\mu} \delta_3 , \nonumber  
\end{align*}
where
\begin{equation}
\delta_e\equiv \frac{e_d}{e_r} Z_1^{\frac{1}{2}}Z_2 - 1,  \qquad
\delta_g\equiv \frac{e_d}{e_r} Z_1^{\frac{1}{2}} Z_2 [\frac{g_d}{g_r}-1] , 
\qquad \delta_3\equiv  \frac{e_d^2}{e_r^2} Z_1 Z_2-1,
\label{ctc}
\end{equation}
and we used the space-time tensor written in terms of the renormalized constant $g_r$ 
\begin{equation}
T^{\mu\nu}_r= g^{\mu\nu}- ig_r M^{\mu\nu}. 
\end{equation}

\begin{figure}[ht]
\centering
\begin{tikzpicture} 
%propagators
\draw[fermion] (-6,4)  -- node[above]{$p$}(-4,4) ;
\node at (-5,3) {$i(p^2-m^2) \delta_{Z_2} - i\delta_m $};
\draw[decorate,decoration=snake] (0,4) -- node[above]{$q$}(2,4);
\node at (1,3) {$ -i(g^{\mu\nu} q^2 -q^\mu q^\nu )\delta_{Z_1}  $};
\node at (0,4.2) {$\mu $};
\node at (2,4.2) {$\nu $};
%vertex
\draw[fermion] (-6,-1) node[above]{$p$} --(-5,0) ;
\draw[fermion] (-5,0)  --  (-4,-1) node[above]{$p^{\prime}$};
\draw[decorate,decoration=snake] (-5,1)--(-5,0);
\node at (-5.5,1.0) {$q, \mu $};
\node at (-5,-2) {$-ie \left[V_{\mu}(p,p')
\delta_{e}  +egM_{\mu\nu} (p'-p)^\nu \delta_g  \right]$};
%seagull
\draw[decorate,decoration=snake] (0,1)--(1,0);
\draw[decorate,decoration=snake] (1,0)--(2,1);
\draw[fermion] (0,-1) node[above]{$p$} --(1,0) ;
\draw[fermion] (1,0)  --  (2,-1) node[above]{$p^{\prime}$};
\node at (0.3,1) {$\mu $};
\node at (1.8,1) {$\nu $};
\node at (1,-2) {$ 2ie^{2} g_{\mu \nu } \delta_3 $};
\end{tikzpicture}
\caption{Feynman rules for the counterterms in the QED of second order fermions.}
\label{CT}
\end{figure}
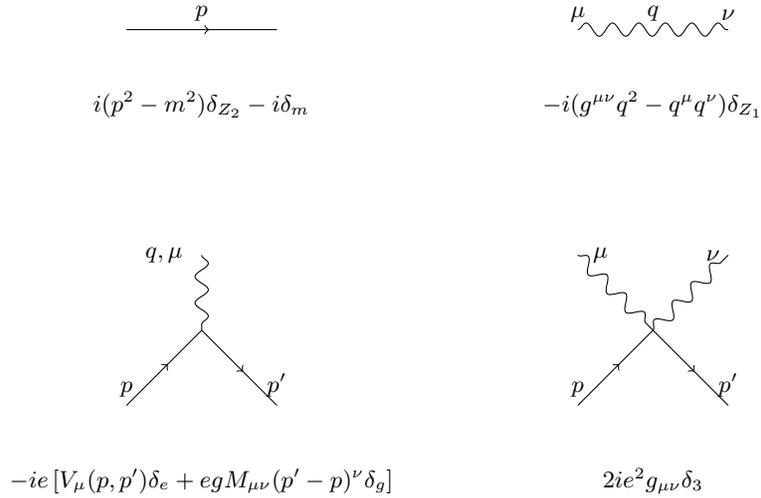

So far we just rewrote the Lagrangian in terms of the renormalized fields and constants 
$m_r,e_r,g_r$. The Feynman rules for the renormalized fields are similar to the ones in 
Fig. \ref{FD} but we now must also include the Feynman rules associated to the generated 
counterterms. These diagrams are shown in Fig. \ref{CT}. Here and in the following, for the sake 
of clarity we will skip the suffix $r$ in the renormalized quantities but will keep the suffix $d$ in 
the bare quantities. 

In the following we use dimensional regularization to handle divergent integrals and carry out 
the renormalization procedure using the mass-shell renormalization conditions.

\subsection{Vacuum polarization}

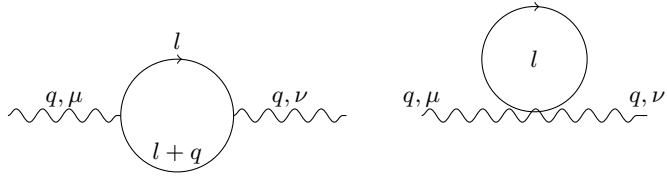
\begin{figure}[ht]
\centering
\begin{tikzpicture} 
\draw[gauge] (-5,0)  -- node[above]{$q,\mu$}(-3.5,0);
\draw[fermionloop] (-2.75,0) circle (.75);
\draw[gauge] (-2,0)  -- node[above]{$q,\nu$}(-.5,0);
\draw[gauge] (.5,0)node[above]{$q,\mu$} -- (3.5,0)node[above]{$q,\nu$};
\draw[fermionloop] (2,.75) circle (.70) ;
\node at (-2.75,1) {$l$};
\node at (-2.75,-.5) {$l+q$};
\node at (2,.75) {$l$};
\end{tikzpicture}
\caption{Feynman diagrams for the vacuum polarization in the QED of second order fermions.} 
\label{polava}
\end{figure}

The vacuum polarization is obtained from Figs. \ref{FD},\ref{CT} as
\begin{equation}
-i\Pi_{\mu\nu}(q)=-i\Pi^{*}_{\mu\nu}(q)-i\delta_{Z_{1}} \left(q^2 g_{\mu\nu}-q_{\mu}q_{\nu}\right),
\label{vacuo}
\end{equation}
where $-i\Pi^{*}_{\mu\nu}(q)$ stands for the contributions from the one-loop diagrams shown in Fig. \ref{polava}.

These diagrams yield the polarization tensor 
\begin{equation}
-i\Pi^{*}_{\mu\nu}(q)=(-\frac{1}{2}) e^2  \mu^{4-d}      \int \frac{d^dl}{(2\pi)^d} 
\left\{  \frac{f(d)(2l+q)_\mu(2l+q)_\nu   +
 \frac{f(d)}{4}  g^2 ( g_{\mu\nu} q^2   - q_\nu q_\mu) }{\square[l+q]\square[l]}  - 
 \frac{2  g_{\mu\nu}}{\square[l]} \right\},
\label{poladef}
\end{equation}
where the $(-\frac{1}{2})$ factor comes from the closed fermion loop, $f(d)=Tr({\bf 1})$ in dimension $d$ with the property $\lim_{d\to 4}f(d)=4$, and we used Eqs. (\ref{trMd}) to calculate 
the trace over the structure of the $(1/2,0)\oplus (0,1/2)$ representation space.  
We use the FeynCalc package \cite{FC} to evaluate the loop integrals and write our results in terms of the 
conventional Passarino-Veltman scalar functions. We obtain the following result for the polarization tensor
\begin{equation}
\Pi^{*\mu\nu}(q)= (q^2g^{\mu\nu}-q^\mu q^\nu) \pi^{*}(q^2),
\end{equation}
where
\begin{equation}
\pi^{*}(q^2) =    \frac{  e^2 } {12\pi^2} \left[\frac{3 g^2-4}{8}
 B_{0}(q^2,m^2,m^2 )  +  \frac{2 m^{2}}{q^2}\left[  B_{0}(q^2,m^2,m^2 )  - B_{0}(0,m^2,m^2 )   \right] 
 -\frac{1}{3} \right],
\end{equation}
with
\begin{equation}
 B_{0}(p_{1}^2,m_{0}^2,m_{1}^2 )=-i  (2 \pi)^{4}\mu^{4-d}      \int \frac{d^dl}{(2\pi)^d} 
 \frac{1 }{(l^{2}-m_{0}^2)((l+p_{1})^2-m_{1}^2)} = B_{0}(p_{1}^2,m_{1}^2,m_{0}^2 ).
\end{equation}
Using  $d= 4-2\epsilon $ and the conventional Feynman parametrization this function can be written as
\begin{equation}
 B_{0}(p_{1}^2,m_{0}^2,m_{1}^2 )=  \frac{1}{\tilde{\epsilon}}+
 \tilde{B}_{0}(p_{1}^2,m_{0}^2,m_{1}^2 ).
\label{B0explicit}
\end{equation}
where
\begin{eqnarray}
\frac{1}{\tilde{\epsilon}} &\equiv& \frac{1}{\epsilon} -\gamma+ \ln{4\pi} \\
 \tilde{B}_{0}(p_{1}^2,m_{0}^2,m_{1}^2 )&\equiv& - \int^1_0 dx 
  \ln \left[\frac{m_{0}^{2}(1-x)+ m_{1}^2 x   - p_{1}^2 x(1-x)}{\mu^{2}} \right].
\label{B0explicit1}
\end{eqnarray}
Some specific values we will need below are
\begin{eqnarray}
B_0(0,m^{2},m^{2})&=&  \frac{1}{\tilde{\epsilon}}-\ln{\frac{m^{2}}{\mu^{2}}}, \nonumber \\
B_0(m^{2},m^{2},0)&=&  \frac{1}{\tilde{\epsilon}}-\ln{\frac{m^{2}}{\mu^{2}}}+2 ,\label{Bs}\\
B_0(0,m^{2},0)&=&  \frac{1}{\tilde{\epsilon}}-\ln{\frac{m^{2}}{\mu^{2}}}+1. \nonumber 
\end{eqnarray}
From Eq. \eqref{vacuo} the complete polarization tensor is given by 
\begin{equation}
\Pi^{\mu\nu}(q)= (q^2g^{\mu\nu}-q^\mu q^\nu) \pi(q^2) ,
\end{equation}
with
\begin{equation}
\pi(q^2)=\pi^{*}(q^2)+\delta_{Z_1}.
\end{equation}
The photon complete propagator is given by the sum of all the 1PI geometric series  
\begin{eqnarray}
i\Delta^{\mu\nu}_c(q)&=& i\Delta^{\mu\nu}(q)+ i\Delta^{\mu\sigma}(q)[-i\Pi_{\sigma \rho}(q)][i \Delta^{\rho\nu}(q)] 
+[i \Delta^{\mu\sigma}(q)] [-i\Pi_{\sigma \rho}(q) ][i\Delta^{\rho\alpha}(q)][-i \Pi_{\alpha\beta}(q)][i \Delta^{\beta\nu}(q)]+...,\nonumber \\
&=&\frac{-g^{\mu\nu}+ q^\mu q^\nu /q^2  }{ [q^2+i\epsilon][1+\pi(q^2)]}.
\label{seriesprop}
\end{eqnarray}

The first renormalization condition  we will use is related to the mass shell condition for the photon,
which in other words requires to prevent the photon to acquire a mass by radiative corrections. This 
imposes the following condition on the polarization form factor
\begin{equation}
\pi(q^2\to 0)=0, 
\label{con1r}
\end{equation}  
which in turn fixes the value of the counterterm as
\begin{align}
\delta_{Z_1}= - \pi^{*}(q^2=0)= - \frac{  e^2 } { 8\pi^2}  
 \left( \frac{g^2}{4}-\frac{1}{3}\right) B_{0}(0,m^2,m^2 )= - \frac{  e^2 } { 8\pi^2} \left( \frac{g^2}{4}-\frac{1}{3}\right)
\left[\frac{1}{\tilde\epsilon}- \ln{\frac{m^2}{\mu^2}}\right] . 
\label{condr11}
\end{align}
Notice that this constant depends on the value of the gyromagnetic factor $g$. The physical form factor is 
then given by
\begin{equation}
\pi(q^2)  =   \frac{  e^2 } {12\pi^2} \left[
\left( \frac{3 g^2-4}{8} +  \frac{2 m^{2}}{q^2}\right)\left[  B_{0}(q^2,m^2,m^2 )  - B_{0}(0,m^2,m^2 )   \right] 
 -\frac{1}{3} \right] .
\end{equation}
Using the explicit representation of $B_0$ in Eqs. (\ref{B0explicit}, \ref{B0explicit1}) we obtain
\begin{equation}
\pi(q^2)  = \frac{ - e^2 } {12\pi^2} \left[
\left( \frac{3 g^2-4}{8} +  \frac{2 m^{2}}{q^2}\right)\left[  \int^1_0 dx 
  \ln{\left(1-\frac{q^2}{m^2}x(1-x) \right) }   \right] 
 +\frac{1}{3} \right].
\end{equation}
In the case $g=2$ we recover the result of Dirac theory. From Eq. \eqref{seriesprop} we see that the 
running of the coupling $\alpha\equiv e^2/4\pi$ induced by the vacuum polarization to this order is 
\begin{align}
\alpha (q^2)=\frac{\alpha (0)}{1+\pi (q^2)}.
\label{run}
\end{align}
In the ultra-relativistic limit $-q^2 \gg m^2$, the vacuum polarization form factor reads
\begin{equation}
\pi(q^{2})|_{-q^2\gg m^2} = \frac{\alpha}{3\pi}\Big[ \frac{3 g^{2}-4}{8}(2- \ln{\frac{-q^2}{m^2}})
-\frac{1}{3}      \Big]. 
\end{equation}
In this limit, the running coupling constant takes the value     
\begin{equation}
\alpha (q^2)|_{-q^2\gg m^2}=  \frac{\alpha(0)}{ 1- \frac{\alpha}{3\pi}\big(1-\frac{3}{2}[1-\frac{g^2}{4}]\big)   
\ln{\frac{-q^2}{Am^2}}},
\end{equation} 
where
\begin{equation}
A\equiv exp\Big\{  \Big(\frac{5}{3}\Big) \frac{1-\frac{9}{5} [1-\frac{g^2}{4}]}{1- \frac{3}{2} [1-\frac{g^2}{4}] }   \Big\}.
\end{equation} 
Notice that the running coupling constant depends in general of the value of $g$ and in the case $g=2$ we recover the 
conventional result of the Dirac theory (see e.g. \cite{peskin} Eq. (7.96)).

\subsection{Fermion self-energy}
\begin{figure}[ht]
\centering
\begin{tikzpicture}
\draw[fermion] (-5,0) node[above]{$p$}  --(-1.5,0) node[above]{$p$};
\draw [decorate,decoration=snake](-2.5,0) arc (0:180:0.75);
\draw[fermion] (.5,0)node[above]{$p$} -- (2,0);
\draw[fermion] (2,0) -- (3.5,0)node[above]{$p$};
\draw[decorate,decoration=snake] (2.0,0.70) circle (.70);
\node at (-3.25,1.25) {$l$};
\node at (-3.25,.25) {$l+p$};
\node at (2,.75) {$l$};
\end{tikzpicture}
\caption{Feynman diagrams for the fermion self-energy.}
\label{self}
\end{figure}
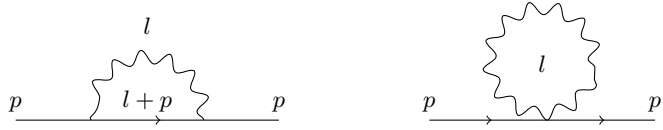

Using the Feynman diagrams in Figs. (\ref{FD},\ref{CT}), the fermion self-energy at one loop level reads
\begin{equation}
-i\Sigma(p^2)= -i \Sigma^{*}(p^2) + i(p^2-m^2) \delta_{Z_2}- i\delta_{m},  \label{renoself}
\end{equation}
where $-i\Sigma^{*}(p^2)$ stands for the one loop diagrams depicted in Fig. \ref{self}.  We use the 
on-shell renormalization condition for this Green function. Similarly to the photon case the complete fermion propagator
is given by
\begin{equation}
S_{c}(p)= \frac{1}{p^2-m^2-\Sigma(p) +i\epsilon}.
\end{equation}
On-shell renormalization requires this function to have a simple pole at $p^2=m^2$ thus the following 
relations must hold
\begin{equation}
\Sigma(p^2=m^2)=0 , \qquad \qquad\frac{\partial \Sigma(p)}{ \partial p^2}\big\vert_{p^2=m^2}=0. \label{con2r}
\end{equation}
These relations fix the counterterms in Eq.(\ref{renoself}) as
\begin{equation}
\delta_m =- \Sigma^{*}(p^2=m^2), \qquad \delta_{Z_2}= \frac{\partial \Sigma^{*}(p^2)}{ \partial p^2}\big\vert_{p^2=m^2},
\end{equation} 
and the renormalized fermion self-energy is given by
\begin{equation}
-i\Sigma(p^2)= -i (\Sigma^{*}(p^2)-\Sigma^{*}(m^2)) + i(p^2-m^2) \frac{\partial \Sigma^{*}(p^2)}{ \partial p^2}\big\vert_{p^2=m^2}.
\label{renoself1}
\end{equation}

%%%%%%
%%%%%
Now we turn to the calculation of diagrams in Fig. (\ref{self}). The tadpole 
diagram vanishes in dimensional regularization. The remaining diagram yields
\begin{equation}
-i\Sigma^{*}(p)= - e^2  \mu^{2\varepsilon}    \int \frac{d^dl}{(2\pi)^d}   \frac{ (2p+l)^2 +
 g^2 M^{\mu\alpha} M_{\mu \beta} l_{\alpha}l^{\beta} }{\square[l+p] \triangle[l]} ,
\end{equation}
with $\triangle[l]\equiv l^{2}-m^{2}_{\gamma}$. In the following we will use a non-vanishing 
photon mass $m_{\gamma}$ to regularize the possible infrared divergences but will keep it in our 
results only in the terms needed for this purpose. With the aid of 
Eq.\eqref{MM-} it is easy to show that  
\begin{equation}
M^{\mu\alpha} M_{\mu\beta} l_\alpha l^\beta = \frac{1}{4}(d -1)l^2   \label{MMll}.
\end{equation}
In terms of the Passarino-Veltman scalar integrals the loop contributions to the fermion self-energy reads
\begin{equation}
\Sigma^{*}(p^2)= \frac{e^2 }{(8\pi)^2}\left[\left( p^{2}+m^{2}\right) B_{0}(p^{2},m^{2},m^{2}_{\gamma}) 
+ \frac{3 g^{2}-4}{8} m^{2}B_{0}(0,m^{2},m^{2})+\frac{g^{2}-4}{8} m^{2} \right].  
 \label{auto}
\end{equation} 
The counterterms in Eq.(\ref{renoself}) are then given by 
\begin{eqnarray}
\delta_m &=&-  \frac{e^2m^2}{(4\pi)^2}
\left[3\left(\frac{g^2}{4}+1\right)\left(\frac{1}{\tilde\epsilon} -\ln{\frac{m^2}{\mu^2}}\right) +
 \frac{g^2}{4}+7 \right],
\label{deltam} \\
\delta_{Z_2}&=&   
\frac{e^2}{8\pi^2}\left[  \frac{1}{\tilde\epsilon}-\ln{\frac{m^2}{\mu^2}} -  \ln{\frac{m^{2}_{\gamma}}{m^2}}\right].
\label{delta2}
\end{eqnarray}
Notice that the renormalization constant of the fermionic field, $Z_2$, does not depend on the 
gyromagnetic factor. There is also an infrared divergence in this constant which we regularize 
with a small photon mass. Finally using Eqs. (\ref{deltam},\ref{delta2}) in Eq. (\ref{renoself}) we get 
the renormalized fermion self-energy as
\begin{equation}
\Sigma(p^2)= \frac{\alpha }{2\pi} \left[
(p^2+m^{2}) \left( B_{0}(p^{2},m^{2},m^{2}_{\gamma})-B_0(m^{2},m^{2},m^{2}_{\gamma}) \right) + 2(p^2-m^{2})
+(p^2-m^{2})  \ln{\frac{m^{2}_{\gamma}}{m^2}} \right].
\end{equation}
Interestingly, the $g$-dependence of this Green function goes away upon renormalization.

\subsection{Fermion-fermion-photon vertex.}
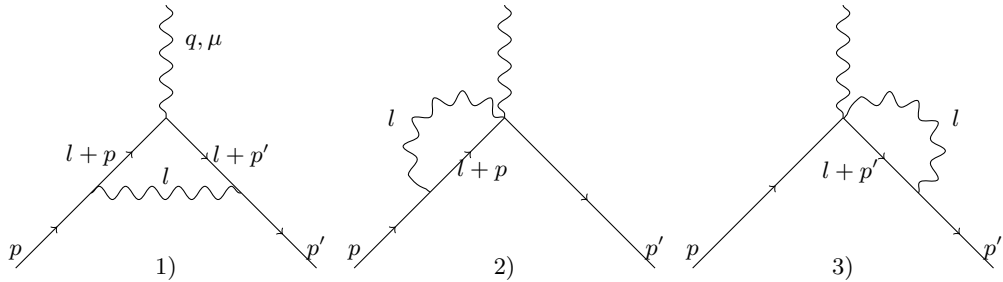
\begin{figure}[ht]
\centering
\begin{tikzpicture} 
%vertex
\draw[fermion] (-5,-2) node[above]{$p$} --(-4,-1.) ;
\draw[fermion] (-4,-1)  --(-3.0,0) ;
\node at (-4.0,-0.5){$l+p$};
\draw[fermion] (-3.0,0)  -- (-2,-1) ;
\node at (-2.0,-0.5){$l+p^{\prime}$};
\draw[fermion] (-2.,-1)  --  (-1,-2) node[above]{$p^{\prime}$};
\draw [decorate,decoration=snake](-4,-1) -- node[above]{$l$}(-2,-1);
\draw[decorate,decoration=snake] (-3,1.5)--(-3,0);
\node at (-2.5,1.0) {$q, \mu $};
\node at (-3,-2) {$1)$};
%vertex2
\draw[fermion] (-0.5,-2) node[above]{$p$} --(0.5,-1.) ;
\draw[fermion] (0.5,-1)  --(1.5,0) ;
\node at (0,0){$l$};
\node at (1.2,-0.7){$l+p$};
\draw[fermion] (1.5,0)  -- (3.5,-2) node[above]{$p^{\prime}$} ;
%\node at (-2.0,-0.5){$l+p^{\prime}$};
\draw[decorate,decoration=snake] (1.5,1.5)--(1.5,0);
%\node at (-2.5,1.0) {$q, \mu $};
\draw [decorate,decoration=snake](1.5,0) arc (45:225:0.7);
\node at (1.5,-2) {$2) $};
%vertex3
\draw[fermion] (4,-2) node[above]{$p$} --(6,0) ;
\draw[fermion] (6,0)  -- (7,-1) ;
\node at (6.1,-0.7){$l+p^{\prime}$};
\node at (7.5,0){$l$};
\draw[fermion] (7,-1)  --(8,-2) node[above]{$p^{\prime}$};
\draw[decorate,decoration=snake] (6,1.5)--(6,0);
%\node at (-2.5,1.0) {$q, \mu $};
\draw [decorate,decoration=snake](7,-1) arc (-45:135:0.7);
\node at (6,-2) {$3) $};
\end{tikzpicture}
\caption{Feynman diagrams for the one loop contributions to the $\gamma ff$ vertex function in the QED of second order fermions.}
\label{ver3lazo}
\end{figure}

The Feynman diagrams in Figs. \ref{FD},\ref{CT} yield the $\gamma ff$ vertex function at one loop level as
\begin{align}
-ie \Gamma^{\mu}(p,q,p')=-ie V^{\mu}(p,p') -ie\Gamma^{*\mu}(p,q,p')
-ie V^{\mu}(p,p')\delta_{e}  -ie[ igM^{\mu\nu} q_\nu] \delta_g   ,
\label{gamareno}
\end{align}
where $\Gamma^{*\mu}(p,q,p')$ stands for the contributions from the one loop diagrams in Fig. \ref{ver3lazo}. 

It can be shown that the one-loop contributions satisfy
\begin{equation}
q^{\mu }\Gamma^{*}_{\mu }(p,q,p^{\prime })=-\Sigma^{*} (p^{\prime 2})+\Sigma^{*} (p^{2}).
\label{wi32bare}
\end{equation}%
Writing this equation in its differential form
\begin{equation}
\Gamma^{*\mu }(p,0,p)=-\frac{\partial \Sigma^{*} (p^{2})}{\partial p_{\mu}},
\label{WI32U}
\end{equation}
and using Eqs.(\ref{1iwd},\ref{gamareno}) we get 
\begin{equation}
\delta_{Z_2}= \delta_e=  Z_1^\frac{1}{2} Z_2 e_d/ e -1,
\label{dz2de}
\end{equation} 
thus the bare and renormalized charges are related as 
\begin{equation}
e= \sqrt{Z_1} e_d. \label{renocarga}
\end{equation}
The one-loop renormalized charge depends only on the renormalization constant for the photon field. Notice that 
this relation fixes also the counterterm for the $\gamma\gamma ff$ vertex function. Indeed, from 
Eqs. (\ref{ctc}) we get 
\begin{equation}
\delta_3=\delta_{Z_2}=\delta_{e}=\frac{e^2}{8\pi^2}\left[  \frac{1}{\tilde\epsilon}-\ln{\frac{m^2}{\mu^2}} -  \ln{\frac{m^{2}_{\gamma}}{m^2}}\right] .
\label{d3}
\end{equation}

For the sake of clarity in the physical interpretation of the different terms arising from the calculation of 
diagrams in Fig. \ref{ver3lazo} we write this vertex function in terms of 
$r\equiv p'+p$ and $q\equiv p'-p$.  The loop contributions to the $\gamma ff$ vertex function are
\begin{equation}
\Gamma^{*\mu}(p,q,p^{\prime})=\mathbb{E}^{*} q^{\mu}+\mathbb{F}^{*} r^{\mu
}+\mathbb{G}^{*}ig M^{\mu\nu}q_{\nu}+\mathbb{H}^{*} igM^{\mu\nu}r_{\nu}%
+\mathbb{I}^{*} igM^{\alpha\beta}r_{\alpha}q_{\beta}\,r^{\mu}+\mathbb{J}^{*}%
igM^{\alpha\beta}r_{\alpha}q_{\beta}\,q^{\mu},
\end{equation}
where $\mathbb{E}^{*}$-$\mathbb{J}^{*}$ are scalar form factors. We write these form factors
in terms of the Passarino-Veltman scalar integrals. A convenient decomposition of the form factors is the following 
\begin{equation}
\mathbb{{\cal F}^{*}}(p^{2},p^{\prime2},q^{2})=
{\displaystyle\sum\limits_{i=0}^{5}}
{\cal F}_{i}\,P_{i}(p^{2},p^{\prime2},q^{2},m^{2},m_{\gamma}^{2},m^{2}),
\label{FFdecomposition}
\end{equation}
where $\mathbb{{\cal F}^{*}=\mathbb{E}^{*} ,\mathbb{F}^{*},\mathbb{G}^{*},\mathbb{H}^{*}
,\mathbb{I}^{*},\mathbb{J}^{*} }$;  ${\cal F}_{i}$ , $i=0,...,5$ are scalar functions, $P_{0}=1$ and $P_{i}$ for
$i=1,...,5$ denote the following Passarino-Veltman scalar integrals:%
\begin{align*}
P_{1} &  =C_{0}(p^{2},p^{\prime2},q^{2},m^{2},m_{\gamma}^{2},m^{2}) ,\\
P_{2} &  =B_{0}(q^{2},m^{2},m^{2}) , \\
P_{3} &  =B_{0}(p^{2},m^{2},0) ,\\
P_{4} &  =B_{0}(p^{\prime2},m^{2},0), \\
P_{5} &  =B_{0}(0,m^{2},0).
\end{align*}
The $C_{0}$ function is given by 
\begin{equation}
C_{0}(p^{2},p^{\prime2},q^{2}%
,m^{2},m_{\gamma}^{2},m^{2})  = -i\left(  4\pi\right)  ^{2} \mu^{4-d}\int\frac{d^{d}l}{\left(
2\pi\right)  ^{d}}\frac{1}{\left(  l^{2}-m_{\gamma}^{2}\right)  \left(
\left(  l+p\right)  ^{2}+m^{2}\right)  \left(  \left(  l+p^{\prime}\right)
^{2}+m^{2}\right)  }.
\end{equation}
The explicit form of the scalar functions ${\cal F}_{i}$  are deferred to the appendix.

The ultraviolet divergences are contained  in the $B_{0}$ functions and are of the form $1/\tilde{\epsilon}$.  
A straightforward calculation yields
\begin{equation}
\sum_{i=2}^{5}E_{i}(p^{2},p^{\prime2},q^{2})=\sum_{i=2}^{5}H_{i}(p^{2},p^{\prime2},q^{2})
=\sum_{i=2}^{5}I_{i}(p^{2},p^{\prime2},q^{2})=\sum_{i=2}^{5}J_{i}(p^{2},p^{\prime2},q^{2})=0, 
\end{equation}
thus the form factors $\mathbb{E}^{*}, \mathbb{H}^{*}, \mathbb{I}^{*}, \mathbb{J}^{*}$ are finite. 
For the charge and magnetic moment form factors we obtain
\begin{eqnarray}
\sum_{i=2}^{5}F_{i}(p^{2},p^{\prime2},q^{2})&=&\frac{-2e^{2}}{(4\pi)^{2}} , \\
\sum_{i=2}^{5}G_{i}(p^{2},p^{\prime2},q^{2})&=&\frac{-e^{2}}{(4\pi)^{2}}\left( \frac{g^{2}}{4}+1 \right).
\end{eqnarray}
These form factors are ultraviolet divergent and need to be renormalized. It is natural to expect the 
divergence of the magnetic moment form factor in our theory since 
here $g$ is a free parameter in the Lagrangian and, on general grounds, it is expected to be renormalized. 

We use on-shell renormalization for the $\gamma ff$ vertex function. Evaluating the scalar form factors at 
$p^2=p'^2=m^2$ and  $q^2=(p'-p)^2=0$  we get
\begin{align}
&\mathbb{F}^{*}_{OS}\equiv \mathbb{F}^{*}(m^{2},m^{2},0)= \frac{2e^2}{(4\pi)^2} \left[2m^{2}   
C_{0}(m^{2},m^{2},0,m^{2},m_{\gamma}^{2},m^{2})-B_{0}(0,m^{2},m^{2}) \right], \\
&\mathbb{G}^{*}_{OS}\equiv \mathbb{G}^{*}(m^{2},m^{2},0) =\mathbb{F}^{*}_{OS}
+ \frac{e^2}{(4\pi)^2} \left[ - B_{0}(0,m^{2},0)  
+ 2 B_{0}(m^{2},m^{2},0)- \frac{g^{2}}{4}B_{0}(0,m^{2},m^{2})-1 \right] , \\
&\mathbb{I}^{*}_{OS}\equiv \mathbb{I}^{*}(m^{2},m^{2},0)  = -\frac{e^2}{(4\pi)^2m^{2}} ,
\end{align}
the remaining form factors vanishing at this kinematical point. Using
\begin{equation}
C_{0}(m^{2},m^{2},0,m^{2},m_{\gamma}^{2},m^{2}) =\frac{1}{2m^{2}} \ln{\frac{m_{\gamma}^{2}}{m^{2}}} ,
\end{equation} 
and the specific values of $B_{0}$ in Eqs.(\ref{Bs}) we obtain
\begin{align}
&\mathbb{F}^{*}_{OS}= \frac{-2e^2}{(4\pi)^2} \left[ \frac{1}{\tilde{\epsilon}}-\ln{\frac{m^{2}}{\mu^{2}}}  
 - \ln{\frac{m_{\gamma}^{2}}{m^{2}}} \right] , \\
&\mathbb{G}^{*}_{OS}=\mathbb{F}^{*}_{OS}+ \frac{e^2}{(4\pi)^2} \left[ \left(
\frac{1}{\tilde{\epsilon}}-\ln{\frac{m^{2}}{\mu^{2}}}\right)\left(1-\frac{g^{2}}{4}\right)+2
 \right].
\end{align}
The on-shell renormalized vertex function in Eq.(\ref{gamareno}) reads
\begin{equation}
-ie \Gamma^{\mu}_{OS}=-ie \left( 1+\delta_{e}+\mathbb{F}^{*}_{OS}\right) r^\mu  -ie \left( 1+\delta_{e} 
+ \delta_g + \mathbb{G}^{*}_{OS}\right) ig M^{\mu\nu} q_\nu 
+\mathbb{I}^{*}_{OS} igM^{ \alpha\beta} r_\alpha q_\beta r^\mu .
\label{gamarenoos}
\end{equation}

The first term defines the physical charge at $q^2=0$. This is the coupling $e$ appearing in our tree level Lagrangian thus
\begin{equation}
\delta_e= -\mathbb{F}^{*}_{OS}=  \frac{e^2}{8\pi^2} \left[ \frac{1}{\tilde{\epsilon}}-\ln{\frac{m^{2}}{\mu^{2}}}   - \ln{\frac{m_{\gamma}^{2}}{m^{2}}} \right] .
\label{deltae2}
\end{equation}
Notice that this is exactly the result in Eq. (\ref{d3}), which we got using the diagrams 
in Fig. \ref{ver3lazo} and the Ward-Takahashi identity in Eq. (\ref{1iwd}). This result for $\delta_{e}$ also cancels one of the 
divergences of the magnetic form factor. In fact, the coefficient of the $ eg M^{\mu\nu} q_\nu $ term in Eq.(\ref{gamarenoos}) reads
\begin{equation}
1+\delta_{e} + \delta_g + \mathbb{G}^{*}_{OS}= 1+ \delta_{g} 
+ \frac{e^2}{(4\pi)^2} \left[ \left(
\frac{1}{\tilde{\epsilon}}-\ln{\frac{m^{2}}{\mu^{2}}}\right)\left(1-\frac{g^{2}}{4}\right)+2
 \right].
\end{equation}
Notice that the divergence of the magnetic form factor associated to $g$ vanishes for $g=\pm 2$. For other values of $g$ we 
need an additional renormalization condition. Unlike the divergence in the charge form factor which is fixed 
by gauge invariance, the magnetic term is gauge invariant by itself and this symmetry does not constrain the corresponding 
parameter. The renormalization condition essentially fixes the value of the parameter in the Lagrangian at some 
scale ($q^2=0$ in this case). Since the divergence vanishes for $g= 2$, it is natural to fix the counterterm to zero 
in this case, which amounts to choose 
\begin{equation}
\delta_g=- \frac{e^2}{(4\pi)^2}\left(
\frac{1}{\tilde{\epsilon}}-\ln{\frac{m^{2}}{\mu^{2}}}\right)\left(1-\frac{g^{2}}{4}\right).
\label{deltag2} 
\end{equation}
This choice yields the one loop correction to the magnetic moment as 
\begin{equation}
\Delta g =  \frac{g}{2} \frac{ \alpha}{\pi}.
\end{equation}
This correction, which depends on the tree level value of the gyromagnetic factor, coincides with the correction in 
the Dirac theory in the case $g= 2$.

In summary, the renormalized $\gamma ff$ vertex function at one loop level reads 
\begin{eqnarray}
\Gamma^{\mu}=  \mathbb{E} ~q^\mu + \mathbb{F}~ r^\mu 
+ \mathbb{G}  igM^{\mu\nu}~ q_\nu   
+ \mathbb{H}~ igM^{\mu\nu}r_\nu  
+ \mathbb{I}~  igM^{ \alpha\beta} r_\alpha q_\beta r^\mu
+\mathbb{J}~ igM^{ \alpha\beta} r_\alpha q_\beta q^\mu,
\end{eqnarray}
with the finite form factors  
\begin{equation}
\mathbb{E} = \mathbb{E}^{*},\qquad
\mathbb{H}= \mathbb{H}^{*},  \qquad
\mathbb{I} = \mathbb{I}^{*}, \qquad
\mathbb{J} = \mathbb{J}^{*}, \qquad
\end{equation}
given in Eq. \eqref{FFdecomposition}, and the renormalized form factors 
\begin{align}
\mathbb{F}(p^{2},p^{\prime2},q^{2})&= 1+\mathbb{F^{*}}(p^{2},p^{\prime2},q^{2})
-\mathbb{F^{*}}(m^{2},m^{2},0), \\
\mathbb{G}(p^{2},p^{\prime2},q^{2})&= 1+\frac{\alpha}{2\pi} +\mathbb{G^{*}}(p^{2},p^{\prime2},q^{2})
-\mathbb{G^{*}}(m^{2},m^{2},0) .
\end{align}

\subsection{Fermion-fermion-photon-photon vertex}

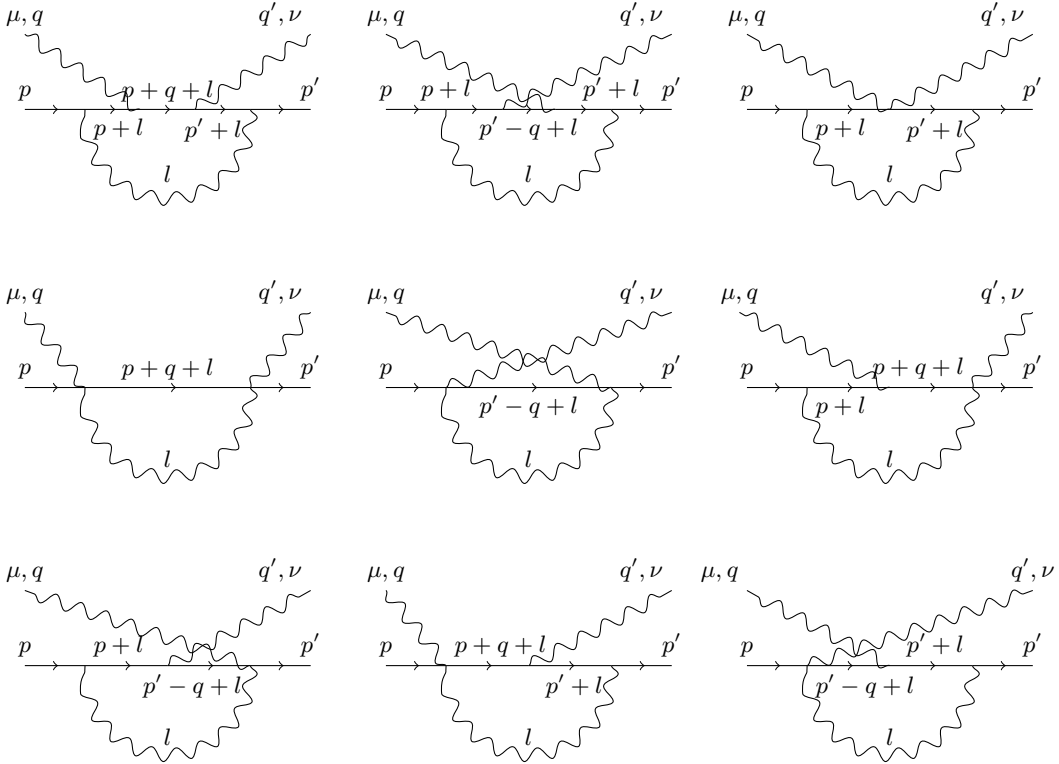
\begin{figure}[ht]
\centering
\begin{tikzpicture} 
%1
\draw[fermion] (-6.7,-1.5)node[above]{$p$}  --(-5.9,-1.5) node[below right]{$p+l$};
\draw[fermion] (-3.7,-1.5)  --(-2.9,-1.5) node[above]{$p'$};
\draw [decorate,decoration=snake](-3.7,-1.5) arc (0:-180:1.1);
\node at (-4.8,- 3.2) {};
\node at (-4.8,- 2.35) {$ l $};
\node at (-4.8, -1.25) {$ p+q+l $};
\draw[fermion] (-5.9,-1.5) --(-5.17,-1.5);
\draw[fermion] (-5.17,-1.5) --(-4.44,-1.5) ;
\draw[fermion] (-4.44,-1.5) --(-3.7,-1.5)node[below left]{$p'+l$} ;
\draw[decorate,decoration=snake] (-4.44,-1.5) --(-2.9,-0.5) node[above left]{$q',\nu$};
\draw[decorate,decoration=snake] (-5.17,-1.5) --(-6.7,-0.5) node[above]{$\mu, q$};
%2
\draw[fermion] (-1.9,-1.5) node[above]{$p$}  --(-1.1,-1.5)node[above]{$p+l$} ;
\draw[fermion] (1.1,-1.5)  --(1.9,-1.5) node[above]{$p'$};
\draw [decorate,decoration=snake](1.1,-1.5) arc (0:-180:1.1);
\node at (0,- 3.2) {};
\node at (0,- 2.35) {$ l $};
\node at (0, -1.75) {$ p'-q+l $};
\draw[fermion] (-1.1,-1.5) --(-0.35,-1.5) ;
\draw[fermion] (-0.35,-1.5) --(0.35,-1.5) ;
\draw[fermion] (0.35,-1.5) --(1.1,-1.5) node[above]{$p'+l$};
\draw[decorate,decoration=snake] (0.35 ,-1.5) --(-1.9,-0.5) node[above]{$\mu, q$};
\draw[decorate,decoration=snake] (-0.35,-1.5) --(1.9,-0.5) node[above left]{$q' , \nu$};
%3
\draw[fermion] (2.9,-1.5) node[above]{$p$}  --(3.7,-1.5) ;
\draw[fermion] (5.9,-1.5)  --(6.7,-1.5) node[above]{$p'$};
\draw [decorate,decoration=snake](5.9,-1.5) arc (0:-180:1.1);
\draw[fermion] (3.7,-1.5)node[below right]{$p+l$} --(4.8,-1.5) ;
\draw[fermion] (4.8,-1.5) --(5.9,-1.5)node[below left]{$p'+l$} ;
\draw[decorate,decoration=snake] (4.8 ,-1.5) --(2.9,-0.5) node[above]{$\mu, q$};
\draw[decorate,decoration=snake] (4.8 ,-1.5) --(6.7,-0.5) node[above left]{$ q',\nu$};
\node at (4.8,- 3.2) {};
\node at (4.8,- 2.35) {$ l $};
%4
\draw[fermion] (-6.7,-5.2)node[above]{$p$}  --(-5.9,-5.2) ;
\draw[fermion] (-3.7,-5.2)  --(-2.9,-5.2) node[above]{$p'$};
\draw [decorate,decoration=snake](-3.7,-5.2) arc (0:-180:1.1);
\node at (-4.8,- 7) {};
\node at (-4.8,- 6.15) {$ l $};
\node at (-4.8,- 4.95) {$ p+q+l $};
\draw[fermion] (-5.9,-5.2) --(-3.7,-5.2) ;
\draw[decorate,decoration=snake] (-5.9,-5.2) --(-6.7,-4.2) node[above]{$\mu,q $};
\draw[decorate,decoration=snake] (-3.7,-5.2) --(-2.9,-4.2) node[above left]{$q',\nu$};
%5
\draw[fermion] (-1.9,-5.2) node[above]{$p$}  --(-1.1,-5.2) ;
\draw[fermion] (1.1,-5.2)  --(1.9,-5.2) node[above]{$p'$};
\draw [decorate,decoration=snake](1.1,-5.2) arc (0:-180:1.1);
\node at (0,- 7) {};
\node at (0,- 6.15) {$ l $};
\node at (0,-5.45) {$ p'-q+l $};
\draw[fermion] (-1.1,-5.2) --(1.1,-5.2) ;
\draw[decorate,decoration=snake] (1.1,-5.2) --(-1.9,-4.2) node[above]{$\mu,q $};
\draw[decorate,decoration=snake] (-1.1,-5.2) --(1.9,-4.2) node[above left]{$q',\nu$};
%6
\draw[fermion] (2.9,-5.2) node[above]{$p$}  --(3.7,-5.2) ;
\draw[fermion] (5.9,-5.2)  --(6.7,-5.2) node[above]{$p'$};
\draw [decorate,decoration=snake](5.9,-5.2) arc (0:-180:1.1);
\node at (4.8,- 7) {};
\node at (4.8,- 6.15) {$ l $};
\draw[fermion] (3.7,-5.2)node[below right]{$p+l$} --(4.8,-5.2) ;
\draw[fermion] (4.8,-5.2) --(5.9,-5.2) node[above left]{$p+q+l$};
\draw[decorate,decoration=snake] (4.8,-5.2) --(2.8,-4.2) node[above]{$\mu,q $};
\draw[decorate,decoration=snake] (5.9,-5.2) --(6.7,-4.2)node[above left]{$q',\nu$};
%7
\draw[fermion] (-6.7,-8.9)node[above]{$p$}  --(-5.9,-8.9) ;
\draw[fermion] (-3.7,-8.9)  --(-2.9,-8.9) node[above]{$p'$};
\draw [decorate,decoration=snake](-3.7,-8.9) arc (0:-180:1.1);
\node at (-4.8,- 10.8) {};
\node at (-4.8,- 9.85) {$ l $};
\draw[fermion] (-5.9,-8.9)node[above right]{$p+l$} --(-4.8,-8.9) ;
\draw[fermion] (-4.8,-8.9) -- (-3.7,-8.9) node[below left]{$p'-q+l$};
\draw[decorate,decoration=snake] (-4.8,-8.9) --(-2.9,-7.9) node[above left]{$q',\nu $};
\draw[decorate,decoration=snake] (-3.7,-8.9) --(-6.7,-7.9) node[above]{$\mu, q$};
%8
\draw[fermion] (-1.9,-8.9) node[above]{$p$}  --(-1.1,-8.9)node[above right]{$p+q+l$} ;
\draw[fermion]  (1.1,-8.9)  --(1.9,-8.9) node[above]{$p'$};
\draw [decorate,decoration=snake](1.1,-8.9) arc (0:-180:1.1);
\node at (0,- 10.8) {};
\node at (0,- 9.85) {$ l $};
\draw[fermion] (-1.1,-8.9) --(0,-8.9) ;
\draw[fermion] (0,-8.9)--(1.1,-8.9)node[below left]{$p'+l$}  ;
\draw[decorate,decoration=snake] (0,-8.9) --(1.9,-7.9) node[above left]{$q',\nu $};
\draw[decorate,decoration=snake] (-1.1,-8.9) --(-1.9,-7.9) node[above]{$\mu, q$};
%9
\draw[fermion] (2.9,-8.9) node[above]{$p$}  --(3.7,-8.9) node[below right]{$p'-q+l$};
\draw[fermion] (5.9,-8.9)  --(6.7,-8.9) node[above]{$p'$};
\draw [decorate,decoration=snake](5.9,-8.9) arc (0:-180:1.1);
\node at (4.8,- 10.8) {};
\node at (4.8,- 9.85) {$ l $};
\draw[fermion] (3.7,-8.9)--(4.8,-8.9) ;
\draw[fermion] (4.8,-8.9)--(5.9,-8.9) node[above left]{$p'+l$};
\draw[decorate,decoration=snake] (4.8,-8.9) --(2.9,-7.9) node[above left]{$ \mu, q $};
\draw[decorate,decoration=snake] (3.7,-8.9) --(6.7,-7.9) node[above]{$q',\nu $};
\end{tikzpicture}
\caption{Feynman diagrams for the $\gamma\gamma ff$ vertex in the QED of second order fermions.}
\label{compton}
\end{figure}

The $\gamma\gamma ff$ vertex function at one loop level is obtained from the Feynman rules in 
Eqs.(\ref{FD},\ref{CT}) as 
\begin{align}
ie^2 \Gamma^{\mu\nu}(p,q,p',q') = ie^2 V^{\mu\nu}(p,q,p',q')+ ie^2 \Gamma^{*\mu\nu}(p,q,p',q') + 2ie^2 g^{\mu \nu } \delta_{3}, 
\label{segwar1}
\end{align}
where the one loop corrections $ie^2 \Gamma^{*\mu\nu}(p,q,p',q')$ are given by the diagrams in Fig. \ref{compton}. 
The couterterm $\delta_{3}$ has been already fixed in Eq. (\ref{d3}) and we must check that this 
counterterm removes all the divergences of these loop diagrams.

It can be shown that the one-loop contributions in Fig. \ref{compton} satisfy 
\begin{align}
k_{\mu}  \Gamma^{*\mu\nu}(p,q,p',q') =   [  \Gamma^{*\nu}(p+q,q',p')-   \Gamma^{*\nu}(p,q',p'-q)   ].
\end{align}
This is the second Ward-Takahashi identity for the one-loop contributions to the $\gamma ff$ and
 $\gamma \gamma ff$ vertex functions . As a cross-check, this relation and the second Ward-Takahashi 
identity in Eq.(\ref{2iwd}) can be used to show that the relation $\delta_{3}=\delta_{e}$ holds.

The divergent pieces of the loop contributions to the $\Gamma^{*\mu\nu}(p,q,p',q')$ vertex function can be isolated 
taking the zero external momentum limit. In this limit, the sum of the first two diagrams can be written as
\begin{align}
i \Gamma^{*\mu\nu}_{1+2} |_{div}= &-e^2\mu^{4-d} \int \frac{d^dl}{(2\pi)^d} 
 \frac{  V^{\alpha}(l,0) \left[ V^\nu(l,l)V^\mu(l,l)+V^\mu(l,l)V^\nu(l,l)\right] V_{\alpha}(0,l)}
 { \square[l]^{3} \triangle[l]^{2}} , \\
 = &-e^2\mu^{4-d} \int \frac{d^dl}{(2\pi)^d} 
 \frac{  8 l^\mu l^\nu \left[l^2+M^{\alpha\beta}M_{\alpha\tau}l^\tau l_\beta  \right] }
 { \square[l]^{3} \triangle[l]} .
\end{align}
Using Eq. (\ref{MMll}) we identify the divergent part as 
\begin{equation}
i\Gamma_{1+2}^{*\mu\nu}|_{div}=-\frac{ie^2}{(4\pi)^2} 2g^{\mu\nu}[1+\frac{3g^2}{4}]
 \frac{1}{\tilde{\epsilon}}. \label{cont1}
\end{equation}
Similarly, the divergent piece of the third diagram is 
\begin{equation}
i \Gamma^{*\mu\nu}_3|_{div} = e^2 \mu^{4-d}  \int \frac{d^dl}{(2\pi)^d} 
\frac{V^\alpha(l,0) 2 g^{\mu\nu} V_{\alpha}(0,l) }{\square[l]^{2}\triangle[l] }
=\frac{ie^2}{(4\pi)^2} 2g^{\mu\nu} (1 + \frac{3 g^2}{4})   \frac{1}{\tilde{\epsilon}}.
\label{cont2}
\end{equation} 
Notice that this divergence cancel the one coming from the first two diagrams in Eq. (\ref{cont1}) yielding
 a finite contribution of the first three diagrams.  The calculation of the next two contributions is straightforward, we obtain
\begin{align}
i\Gamma^{*\mu\nu}_{4+5}|_{div}=  
-\frac{ie^2}{(4\pi)^{2}} 8 g^{\mu\nu} \frac{1}{\tilde{\epsilon}}.
\label{cont3}
\end{align}
In a similar way, in the zero external momentum limit the sum of the remaining diagrams yield
\begin{equation}
 i\Gamma^{*\mu\nu}_{6+7+8+9}|_{div} = ie^2 \int \frac{d^dl}{(2\pi)^d} 
 \frac{ 16l^{\mu}l^\nu} {\square[l]^{2}\triangle[l]  } 
 = \frac{ie^2}{(4\pi)^2}4 g^{\mu\nu} \frac{1}{\tilde{\epsilon}} .
\label{cont4}
\end{equation}
Finally, adding up the contributions in Eqs. (\ref{cont1},\ref{cont2},\ref{cont3},\ref{cont4}), 
we obtain the divergent part of the loop contributions to the $\gamma\gamma ff$ vertex function as 
\begin{equation}
i\Gamma^{*\mu\nu}(p,k,p',k')|_{div}=  
-2 \frac{ie^2}{(4\pi)^2} [2 g^{\mu\nu}] \frac{1}{\tilde{\epsilon}}.
\label{ffppinf}
\end{equation}
The divergent part is proportional 
to $g^{\mu\nu}$ and using the value of $\delta_{3}$ (Eq.(\ref{d3})) in Eq. (\ref{segwar1}), we obtain 
\begin{align}
ie^2 \Gamma^{*\mu\nu}|_{div} + 2ie^2 g^{\mu \nu } \delta_{3}=0, 
\end{align}
thus the renormalized $\gamma\gamma ff$ vertex function in Eq. \eqref{segwar1} is free of ultraviolet divergences. 

This completes the one loop calculation of the divergences of the renormalized vertex functions appearing 
in the Lagrangian for the quantum electrodynamics of second order fermions in the Poincar\'e projector 
formalism. All these vertex functions are free of ultraviolet divergences to this order.
From our power counting analysis of the superficial degree of the divergence of vertex functions, only those 
with at most four external legs can be divergent. The complete proof of the renormalizability of the 
formalism requires the analysis of divergences of the $3\gamma$,$4\gamma$ and $4f$ vertex functions. The $3\gamma$ 
vertex function must vanish because of charge conjugation symmetry. We will analyze the remaining two vertex 
functions and the physics of the calculated form factors in a future work.

\section{Conclusions} 
In this work we analyze the superficial degree of divergence of the vertex functions of the electrodynamics 
of fermions in the Poincar\'e projector formalism which is second order in the derivatives of the fields. We 
calculate at one loop level the vacuum polarization, the fermion self-energy and the $\gamma-fermion-fermion$ 
vertex functions. We also calculate the divergent part of the one-loop contributions to the 
$\gamma-\gamma-fermion-fermion$ vertex function and show that it is renormalizable.  We obtain a photon 
propagator that depends on the tree level value of $g$ which yields a $g$-dependence of 
the running coupling constant $\alpha (q^2)$. The fermion self-energy turns out to be independent of $g$. 
In addition to the conventional divergence related to the charge form factor, the one-loop contributions to the 
magnetic moment form factor contain a divergence associated to the gyromagnetic factor which vanishes for $g=\pm 2$. 
This requires the gyromagnetic factor to be renormalized in the general case  and in this sense is a true coupling 
running with the energy. As we do with every coupling in the Lagrangian, we must fix the value of $g(q^2)$ at some 
energy scale. Since the divergence vanishes for $g=2$ it is natural to choose the corresponding counterterm to remove the 
$g$-dependent divergence in such a way that for a particle with $g=2$ there is no need to renormalize this coupling.  
This choice leads to a one-loop correction $\Delta{g}=g\alpha/2\pi$ for the gyromagnetic factor and for $g=2$ we recover 
results of Dirac theory for the photon propagator, the running of $\alpha$ and the one loop corrections to the 
gyromagnetic factor.

\acknowledgments{This work was supported by CONACyT under project 156618 and by DAIP-UG. We thank 
Selim G\'omez-\'Avila and Carlos A. Vaquera-Araujo for useful conversations on the subject of this paper. Special thanks 
are given to C.A.V-A. for a cross check of the calculations presented here. We also thank H. Pietschmann for 
sending us his paper in \cite{hebert} and Wolfgang Bietenholz for a critical reading of the manuscript.  }

\section{Appendix.}
\subsection{Lorentz structure and $d$ dimensional calculus.}

The generators of the Homogeneous Lorentz Group (HLG) are the rotation and boost generators
$\{\mathbf{J},\mathbf{K}\}$ which satisfy the following algebra%
\begin{equation}
\lbrack J_{i},J_{j}]=i\epsilon_{ijk}J_{k},\qquad\lbrack J_{i},K_{j}%
]=i\epsilon_{ijk}K_{k},\qquad\lbrack K_{i},K_{j}]=-i\epsilon_{ijk}J_{k}.
\end{equation}
The part of the HLG connected to the identity is isomorphic to the
$SU(2)_{A}\otimes SU(2)_{B}$ group generated by the operators%
\begin{equation}
\mathbf{A}=\frac{1}{2}(\mathbf{J}-i\mathbf{K}),\qquad\mathbf{B}=\frac{1}%
{2}(\mathbf{J}+i\mathbf{K}), \label{AB}%
\end{equation}
hence the irreducible representations (irrep\rq{}s) of the HLG can be characterized by two independent $SU(2)$
quantum numbers $(a,b)$. A given irrep $(a,b)$ has dimension $(2a+1)(2b+1)$
and the states in this irrep are labeled by the corresponding quantum numbers
$|a,m_{a};b,m_{b}\rangle$ where $m_{a}$ and $m_{b}$ are the eigenvalues of
$A_{3}$ and $B_{3}$ respectively. The irreps with well defined value of
$\mathbf{J}^{2}$ are those with $a=0$ or $b=0$. In the case $b=0$ the
representations of the rotations and boost generators are related as
$\mathbf{J}=-i\mathbf{K}$ and since $\mathbf{A}=\mathbf{J}$ we denote these
irreps as $(j,0)$ and refer to them as {\it right } representations of spin $j$ .
In the case $a=0$ we get $\mathbf{J}=i\mathbf{K}$, thus $\mathbf{B}%
=\mathbf{J}$ and we denote these irreps as $(0,j)$ and refer to them as
{\it left } representations of spin $j$ . Since we know how to construct a
representation for the $SU(2)$ rotation group, in both cases we have a
representation for the boost operator and it is possible to explicitly
construct the states in the basis $|j,m\rangle$ of well defined $\mathbf{J}%
^{2}$ and $J_{3}$ starting with the rest frame states \cite{DN}. Here we are 
just interested in the properties of the generators which will enter our calculations.  
In the case $(\frac{1}{2},0)$ and in the conventional basis
$|\frac{1}{2},m\rangle$ of eigenstates of $\{\mathbf{J}^{2},J_{3}\}$  the
generators of rotations are $\mathbf{J}=\bm{\sigma}/2$ and the generators of the
HLG are
\begin{equation}
M_{R}^{ij}=\varepsilon_{ijk}J_{Rk}=\frac{1}{2}\varepsilon_{ijk}\sigma_{k}%
=\frac{1}{4i}[\sigma_{i},\sigma_{j}],\qquad M_{R}^{0i}=K_{Ri}=iJ_{Ri}=\frac
{i}{2}\sigma_{i}. \label{MijR}%
\end{equation}
Similarly the generators for the $(0,\frac{1}{2})$ representation are%
\begin{equation}
M_{L}^{ij}=\varepsilon_{ijk}J_{Lk}=\frac{1}{2}\varepsilon_{ijk}\sigma_{k}%
=\frac{1}{4i}[\sigma_{i},\sigma_{j}],\qquad M_{L}^{0i}=K_{Li}=-iJ_{Li}%
=-\frac{i}{2}\sigma_{i}. \label{MijL}%
\end{equation}

The description of the interactions of spin $\frac{1}{2}$ particles according
to the gauge principle requires to construct first a Lagrangian for the free
particle. This is a scalar function and it was shown in \cite{DN} that it is not
possible to construct a Lagrangian using only two-dimensional left or right
spinors. This can be done only at the price of enlarging the representation space 
to $(\frac{1}{2},0)\oplus(0,\frac{1}{2})$.
The generators for $(\frac{1}{2},0)\oplus(0,\frac{1}{2})$ read
\begin{equation}
M^{ij}=\varepsilon_{ijk}J_{k}\equiv\frac
{1}{2}\sigma^{ij},\qquad M^{0i}=K_{i}\equiv\frac{1}{2}\sigma^{0i},
\label{bw1}
\end{equation}
where%
\begin{equation}
J_{k}=\frac{1}{2}\left(
\begin{array}
[c]{cc}%
\sigma_{k} & 0\\
0 & \sigma_{k}%
\end{array}
\right)  ,\qquad K_{i}=\frac{i}{2}\left(
\begin{array}
[c]{cc}%
\sigma_{i} & 0\\
0 & -\sigma_{i}%
\end{array}
\right)   .
\label{bw2}
\end{equation}
Notice that these relations define the matrices $\sigma^{\mu\nu}$ in terms of the generators $M^{\mu\nu}$. 
The generators form an anti-symmetric Lorentz tensor and, although we will not use this form in our work, 
it is easy to show that these matrices can be also written in the conventional form
\begin{equation}
\sigma^{\mu\nu}=\frac{i}{2}[\gamma^{\mu},\gamma^{\nu}]
\end{equation}
with
\begin{equation}
\gamma^{i}=\left(
\begin{array}
[c]{cc}%
0 & -\sigma_{i}\\
\sigma_{i} & 0
\end{array}
\right)  ,\qquad\gamma^{0}=\left(
\begin{array}
[c]{cc}%
0 & 1\\
1 & 0
\end{array}
\right).
\end{equation}

Notice that the boost generators can be written as $\mathbf{K}=i\chi\mathbf{J}$ where $\chi$ is the hermitian matrix%
\begin{equation}
\chi=\left(
\begin{array}
[c]{cc}%
1 & 0\\
0 & -1
\end{array}
\right)  .
\end{equation}

The eigenstates of this operator are the chiral {\it left} and {\it right } states
embedded in this larger representation space. Therefore we call it \textit{chirality operator} 
in the following and sticking to the conventional notation we will write it as $\chi = \gamma^{5}$.
The relation $\mathbf{K}=i\chi\mathbf{J}$ 
can be inverted to yield $\chi= -i \frac{4}{3}\mathbf{J}\cdot\mathbf{K}$ which reveals 
this operator as proportional to one of the Casimir operators of the Lorentz group in this 
representation. Indeed, it can be rewritten in terms of the generators as
\begin{equation}
\gamma^5= - \frac{i}{3!} \tilde M^{\mu\nu} M_{\mu\nu} , 
\end{equation}
with $\tilde M^{\mu\nu}= \epsilon^{\alpha\beta\mu\nu}M_{\alpha\beta}$. It is worthy to remark that although 
this equation reveals $\gamma^{5}$ as a proportional to a Casimir operator in the $(1/2,0)\oplus (0,1/2)$ 
representation of the Lorentz group, it is not proportional to the unity operator because this is a reducible 
representation whose irreducible sectors are distinguished precisely by the eigenvalues of this operator. 

In our calculations we need multiple products of the generators. We calculate here the simplest product 
\begin{align}
M^{\alpha\beta} M^{\mu\nu}= \frac{1}{2} [M^{\alpha\beta}, M^{\mu\nu}]+ 
\frac{1}{2}\{ M^{\alpha\beta}, M^{\mu\nu}\}.
\end{align}
The anti-symmetric part obeys the Lorentz commutation rules
\begin{align}
 [M^{\alpha\beta}, M^{\mu\nu}]  = - i ( g^{\alpha\mu} M^{\beta\nu}  - g^{\alpha\nu} M^{\beta\mu}+ 
 g^{\beta\nu} M^{\alpha\mu}  - g^{\beta\mu}M^{\alpha\nu} ). 
\label{MM-}               
\end{align}
The symmetric part can be easily calculated using Eqs. (\ref{bw1},\ref{bw2}). We obtain 
\begin{equation}
\{M^{\mu\nu},M^{\alpha\beta}\}= \frac{1}{2}(g^{\mu\alpha}g^{\nu\beta}-g^{\mu\beta}g^{\nu\alpha})
+\frac{i}{2}\epsilon^{\mu\nu\alpha\beta} \gamma^5 .
\label{MM+}
\end{equation}
Higher products of the generators can be calculated using recursively these relations. 
We also need to calculate the trace of the product of generators. The simplest one is  
\begin{equation}
tr \left( M^{\mu\nu}\right)=0,
\end{equation}
as can be directly verified from Eqs. (\ref{bw1},\ref{bw2}) or derived using Lorentz covariance.
Using this relation and Eqs. (\ref{MM-},\ref{MM+}) we obtain 
\begin{equation}
tr\left( M^{\mu\nu}M^{\alpha\beta} \right)= 
\frac{1}{4}(g^{\mu\alpha}g^{\nu\beta}-g^{\mu\beta}g^{\nu\alpha}) ~tr(\mathbf{1}), 
\label{trMd}
\end{equation}
where we also used 
\begin{equation}
tr(\gamma_{5})=0.
\label{trg5}
\end{equation}

In $d$ dimensions we assume that the generators 
still satisfy the Lorentz algebra in Eq. (\ref{MM-}) and the anti-commutator relation in Eq.(\ref{MM+}), but 
now $g^{\mu}_{\,\,\mu}= d$ and $tr(\mathbf{1})=f(d)$ where $f$ is a smooth function of $d$ with the property 
$\lim_{d\to 4} f(d)=4$. The generators still are traceless and on the light of the interpretation of the 
chirality operator we still require it to satisfy Eq.(\ref{trg5}) in $d$ dimensions.

\subsection{Scalar functions for the decomposition of the form factors of the three point function $\gamma ff$}

The scalar functions ${\cal F}_{i} =E_{i}, F_{i}, G_{i}, H_{i}, J_{i}, I_{i}$  entering the decomposition of the form 
factors in Eq. (\ref{FFdecomposition}) are the following functions:
\subsubsection{ $ E_{i}$}
\begin{align*}
E_{0} &  =0, \\
E_{1} &  =\zeta\left(  p^{2}-p^{\prime 2}\right)  \left[  
\left(  g^{2}-4\right)  \left[  m^{2}q^{2}+p\cdot p^{\prime
}\left(  p^{2}+p^{\prime}{}^{2}\right)  -2p^{2}p^{\prime}{}^{2}\right]
+8\left[  m^{4}+2p\cdot p^{\prime}\left(  m^{2}+\left(  p\cdot p^{\prime
}\right)  \right)  -p^{2}p^{\prime}{}^{2}\right]  \right] , \\
E_{2} &  =-\zeta\left(  p^{2}-p^{\prime}{}^{2}\right)  \left[  \left(
g^{2}-4\right)  q^{2}+8\left(  p\cdot p^{\prime}+m^{2}\right)  \right] , \\
E_{3} &  =\zeta\left[\left(  g^{2}-4\right)  \left(  p^{2}-p^{\prime}{}^{2}\right)
\left(  p^{2}-p\cdot p^{\prime}\right)  +8p^{2}\left(  p^{\prime}{}^{2}%
+m^{2}\right)  +8p\cdot p^{\prime}\left(  p^{2}+m^{2}\right)\right] , \\
E_{4} &  =\zeta \left[\left(  g^{2}-4\right)  \left(  p^{2}-p^{\prime}{}^{2}\right)
\left(  p^{\prime2}-p\cdot p^{\prime}\right)  -8p^{\prime2}\left(  p^{2}%
+m^{2}\right)  -8p\cdot p^{\prime}\left(  p^{\prime2}+m^{2}\right)\right] , \\
E_{5} &  =0. 
\end{align*}

\subsubsection{ $F_{i}$}

\begin{align*}
F_{0} &  =0,\\
F_{1} &  =\zeta q^{2}\left[ \left(  g^{2}-4\right)  \left[  m^{2}q^{2}+p\cdot p^{\prime
}\left(  p^{2}+p^{\prime}{}^{2}\right)  -2p^{2}p^{\prime}{}^{2}\right]
+8\left[  m^{4}+2p\cdot p^{\prime}\left(  m^{2}+\left(  p\cdot p^{\prime
}\right)  \right)  -p^{2}p^{\prime}{}^{2}\right]  \right]  ,\\
F_{2} &  =-\zeta q{}^{2}\left[  \left(  g^{2}-4\right)  q{}^{2}+8\left(  p\cdot
p^{\prime}+m^{2}\right)  \right]  ,\\
F_{3} &  =\zeta\left[\left(  g^{2}-4\right)  \left(  p^{2}-p\cdot p^{\prime}\right)
q{}^{2}+8p\cdot p^{\prime}\left(  p^{2}-m^{2}\right)  -8p{}^{2}\left(
p^{\prime2}-m^{2}\right) \right], \\
F_{4} &  =\zeta\left(  g^{2}-4\right)  \left(  p^{\prime}{}^{2}-p\cdot
p^{\prime}\right)  q^{2}+8p\cdot p^{\prime}\left(  p^{\prime2}-m^{2}\right)
-8p^{\prime}{}^{2}\left(  p^{2}-m^{2}\right) , \\
F_{5} &  =0.
\end{align*}

\subsubsection{ $ G_{i}$}

\begin{align*}
G_{0} &  =0,\\
G_{1} &  =2\zeta \left[ 2m^{4}q{}^{2}+2m^{2}\left(  p^{\prime}{}^{2}-p^{2}\right)  ^{2}+4p\cdot
p^{\prime}\left[  \left(  m^{2}+p\cdot p^{\prime}\right)  q{}^{2}-2\left(
\left(  p\cdot p^{\prime}\right)  ^{2}-p^{2}p^{\prime}{}^{2}\right)  \right] \right. 
\\
&  \left. +2p^{\prime}{}^{4}\left(  p\cdot p^{\prime}-p^{2}\right)  +2p^{4}\left(
p\cdot p^{\prime}-p^{\prime}{}^{2}\right)  +\left(  g-2\right)  \left(
m^{2}+p\cdot p^{\prime}\right)  \left(  p^{\prime}{}^{2}-p^{2}\right)  ^{2} \right] , \\
G_{2} &  =-2\zeta\left[  2\left(  m^{2}+p\cdot p^{\prime}\right)  q^{2}+g\left(
p^{2}-p^{\prime}{}^{2}\right)  ^{2}+\left(  g^{2}+4\right)  \left(
p^{2}p^{\prime}{}^{2}-\left(  p\cdot p^{\prime}\right)  ^{2}\right)  \right], \\
G_{3} & =\frac{2\zeta}{p^{2}}\left[  2p^{2}\left(m^{2} + p\cdot p^{\prime}\right)
\left(  p^{2}-p\cdot p^{\prime}\right)  +2m^{2}\left(  p^{2}p^{\prime}{}%
^{2}-\left(  p\cdot p^{\prime}\right)  ^{2}\right)  +gp^{2}\left(
p^{2}-p^{\prime}{}^{2}\right)  \left(  p^{2}+p\cdot p^{\prime}\right)
\right] , \\
G_{4} & =\frac{2\zeta}{p^{\prime}{}^{2}}\left[  2p^{\prime}{}^{2}\left(  m^{2}+p\cdot
p^{\prime}\right)  \left(  p^{\prime}{}^{2}-  p\cdot p^{\prime}
\right)  +2m^{2}\left(  p^{2}p^{\prime}{}^{2}-\left(  p\cdot p^{\prime
}\right)  ^{2}\right)  +gp^{\prime}{}^{2}\left(  p^{\prime}{}^{2}%
-p^{2}\right)  \left(  p^{\prime}{}^{2}+p\cdot p^{\prime}\right)  \right] , \\
G_{5} &  =\zeta \left[\frac{4m^{2}}{p^{2}p'^{2}}\left(  p^{\prime}{}^{2}+p^{2}\right)  \left(  \left(  p\cdot
p^{\prime}\right)  ^{2}-p^{2}p^{\prime}{}^{2}\right) \right] .
\end{align*}

\subsubsection{ $H_{i}$}

\begin{align*}
H_{0} &  =0, \\
H_{1} &  =2\zeta g\left(  p^{2}-p^{\prime}{}^{2}\right)  \left(  m^{2}%
+p\cdot p^{\prime}\right)  q^{2} , \\
H_{2} &  =-2\zeta g\left(  p^{2}-p^{\prime}{}^{2}\right)  q^{2} , \\
H_{3} &  =\frac{2\zeta}{p^{2}}\left[  -2\left(  p^{2}-m^{2}\right)  \left(  \left(
p\cdot p^{\prime}\right)  ^{2}-p^{2}p^{\prime}{}^{2}\right)  +gp^{2}%
q^{2}\left(  p^{2}+p\cdot p^{\prime}\right)  \right] , \\
H_{4} &  =-\frac{2\zeta}{p^{\prime}{}^{2}}\left[  -2\left(  p^{\prime}{}^{2}%
-m^{2}\right)  \left(  \left(  p\cdot p^{\prime}\right)  ^{2}-p^{2}p^{\prime
}{}^{2}\right)  {}+gp^{\prime}{}^{2}q^{2}\left(  p^{\prime}{}^{2}+p\cdot
p^{\prime}\right)  \right]  ,\\
H_{5} &  =\frac{4\zeta m^{2}\left(  p^{2}-p^{\prime}{}^{2}\right)  }%
{p^{2}p^{\prime}{}^{2}}\left(  \left(  p\cdot p^{\prime}\right)  ^{2}%
-p^{2}p^{\prime}{}^{2}\right).
\end{align*}

\subsubsection{ $I_{i}$}

\begin{align*}
I_{0} &  =2\zeta q^{2} , \\
I_{1} &  =\frac{\zeta}{\left(  \left(  p\cdot p^{\prime}\right)  ^{2}%
-p^{2}p^{\prime}{}^{2}\right)  }\left[  3m^{4}p^{\prime}{}^{4}+6m^{2}%
p^{\prime}{}^{2}\left(  p^{\prime}{}^{2}-2m^{2}\right)  p\cdot p^{\prime
}+\left(  8m^{2}-4p^{\prime}{}^{2}\right)  \left(  p\cdot p^{\prime}\right)
^{3}+p^{6}p^{\prime}{}^{2}\right.  \\
&  \left.  +2\left(  6m^{4}-8m^{2}p^{\prime}{}^{2}+p^{\prime}{}^{4}\right)
\left(  p\cdot p^{\prime}\right)  ^{2}+p^{4}\left(  3m^{4}-8m^{2}p^{\prime}%
{}^{2}+\left(  6m^{2}-8p^{\prime}{}^{2}\right)  p\cdot p^{\prime}+2p^{\prime
}{}^{4}+2\left(  p\cdot p^{\prime}\right)  ^{2}\right)  \right.  \\
&  \left.  +p^{2}\left(  -16\left(  m^{2}-p^{\prime}{}^{2}\right)  \left(
p\cdot p^{\prime}\right)  ^{2}+p^{\prime}{}^{2}\left(  6m^{4}-8m^{2}p^{\prime
}{}^{2}+p^{\prime}{}^{4}\right)  -4\left(  3m^{4}-7m^{2}p^{\prime}{}%
^{2}+2p^{\prime}{}^{4}\right)  p\cdot p^{\prime}-4\left(  p\cdot p^{\prime
}\right)  ^{3}\right)  \right] , \\
I_{2} &  =-\frac{\zeta q^{2}}{\left(  p\cdot p^{\prime}\right)  ^{2}-p^{2}p^{\prime}%
{}^{2}}\left[  3\left(  p^{2}+p^{\prime}{}^{2}\right)  \left(  m^{2}+p\cdot
p^{\prime}\right)  -2\left(  p\cdot p^{\prime}\right)  \left(  p\cdot
p^{\prime}+3m^{2}\right)  -4p^{\prime}{}^{2}p^{2}\right] , \\
I_{3} &  =  \frac{\zeta}{p^{2}\left(  \left(  p\cdot p^{\prime}\right)  ^{2}%
-p^{2}p^{\prime}{}^{2}\right)  }\left[  3p^{6}\left(  m^{2}+p\cdot p^{\prime
}-p^{\prime}{}^{2}\right)  +p^{4}\left(  -9m^{2}p\cdot p^{\prime}+p^{\prime}%
{}^{2}\left(  5m^{2}+7p\cdot p^{\prime}\right)  -6\left(  p\cdot p^{\prime
}\right)  ^{2}-p^{\prime}{}^{4}\right)  \right.  \\
& \left.  +p^{2}\left(  4m^{2}\left(  p\cdot p^{\prime}\right)  ^{2}%
+p^{\prime}{}^{2}\left(  -5m^{2}p\cdot p^{\prime}-2\left(  p\cdot p^{\prime
}\right)  ^{2}\right)  +2\left(  p\cdot p^{\prime}\right)  ^{3}\right)
+2m^{2}\left(  p\cdot p^{\prime}\right)  ^{3}\right] ,
\\
I_{4} &  = \frac{\zeta}{p^{\prime}{}^{2}\left(  \left(  p\cdot p^{\prime}\right)
^{2}-p^{2}p^{\prime}{}^{2}\right)  }\left[  3p^{\prime}{}^{6}\left(
m^{2}+p\cdot p^{\prime}-p^{2}\right)  +p^{\prime}{}^{4}\left(  -9m^{2}p\cdot
p^{\prime}+p^{2}\left(  5m^{2}+7p\cdot p^{\prime}\right)  -6\left(  p\cdot
p^{\prime}\right)  ^{2}-p^{4}\right)  \right.  \\
& \left.  +p^{\prime}{}^{2}\left(  4m^{2}\left(  p\cdot p^{\prime}\right)
^{2}+p^{2}\left(  -5m^{2}p\cdot p^{\prime}-2\left(  p\cdot p^{\prime}\right)
^{2}\right)  +2\left(  p\cdot p^{\prime}\right)  ^{3}\right)  +2m^{2}\left(
p\cdot p^{\prime}\right)  ^{3}\right] , \\
I_{5} &  =-\frac{2\zeta m^{2}}{p^{2}p^{\prime}{}^{2}}\left[  \left(  p^{2}+p^{\prime}%
{}^{2}\right)  p\cdot p^{\prime}-2p^{2}p^{\prime}{}^{2}\right]  .
\end{align*}

\subsubsection{ $ J_{i}$}

\begin{align*}
J_{0} &  =2\zeta\left(  p^{2}-p^{\prime}{}^{2}\right) , \\
%%%%%%%%%%%%%
J_{1} &  = \frac{\zeta\left(  p^{2}-p^{\prime2}\right)  }{\left(  p\cdot p^{\prime
}\right)  ^{2}-p^{2}p^{\prime2}}\left[  2g\left(  p\cdot p^{\prime}%
+m^{2}\right)  \left[  \left(  p\cdot p^{\prime}\right)  ^{2}-p^{2}p^{\prime
}{}^{2}\right]  +q^{2}\left(  6m^{2}p\cdot p^{\prime}+3m^{4}+p^{\prime}{}%
^{2}p^{2}\right)  \right.  \\
& \left.  +8m^{2}\left[  \left(  p\cdot p^{\prime}\right)  ^{2}-p^{\prime}%
{}^{2}p^{2}\right]  +2\left(  p\cdot p^{\prime}\right)  ^{2}\left(
p^{2}+p^{\prime}{}^{2}\right)  -4p^{\prime}{}^{2}p^{2}p\cdot p^{\prime
}\right] ,\\
%%%%%%%%%%%%%%%%
J_{2} &  =\frac{\zeta\left(  p^{2}-p^{\prime2}\right)  }{\left(  p\cdot p^{\prime
}\right)  ^{2}-p^{2}p^{\prime2}}\left[  -2g\left(  \left(  p\cdot p^{\prime
}\right)  ^{2}-p^{2}p^{\prime}{}^{2}\right)  -3m^{2}q^{2}-3p\cdot p\left(
^{\prime}p^{2}+p^{\prime}{}^{2}\right)  +4p^{\prime}{}^{2}p^{2}+2\left(
p\cdot p^{\prime}\right)  ^{2}\right] ,\\
%%%%%%%%%%%%%%%
J_{3} &  = \frac{\zeta}{p^{2}\left(  \left(  p\cdot p^{\prime}\right)  ^{2}%
-p^{2}p^{\prime2}\right)  }\left[  2gp^{2}\left(  \left(  p^{2}+p\cdot
p^{\prime}\right)  \left[  \left(  p\cdot p^{\prime}\right)  ^{2}%
-p^{2}p^{\prime}{}^{2}\right]  \right)  \right.  \\
& \left.  -p^{2}\left[  3p^{4}-p^{2}p^{\prime}{}^{2}-2\left(  p\cdot
p^{\prime}\right)  ^{2}\right]  \left(  p^{\prime}{}^{2}-m^{2}\right)  -p\cdot
p^{\prime}\left[  5p^{\prime}{}^{2}p^{2}-3p^{4}-2\left( p\cdot p^{\prime}\right)^{2} \right]
\left(  p^{2}-m^{2}\right)  \right],\\
%%%%%%%%%%%%%%
J_{4} &  = \frac{\zeta}{p^{\prime2}\left(  \left(  p\cdot p^{\prime}\right)
^{2}-p^{2}p^{\prime2}\right)  }\left[  -2gp^{\prime}{}^{2}\left(  p^{\prime}%
{}^{2}+p\cdot p^{\prime}\right)  \left[  \left(  p\cdot p^{\prime}\right)
^{2}-p^{2}p^{\prime}{}^{2}\right]  \right.  \\
&  \left.  +p^{\prime}{}^{2}\left[  3p^{\prime}{}^{4}-p^{\prime}{}^{2}%
p^{2}-2\left(  p\cdot p^{\prime}\right)  ^{2}\right]  \left(  p^{2}%
-m^{2}\right)  +p\cdot p^{\prime}\left[  5p^{2}p^{\prime}{}^{2}-3p^{\prime}%
{}^{4}-2\left( p\cdot p^{\prime}\right)^{2}\right]  \left(  p^{\prime}{}^{2}-m^{2}\right)
\right] ,  \\
%%%%%%%%%%%%%%%%
J_{5} &  =-\frac{2\zeta m^{2} }{p^{2}p'^{2}} \left(p^{2}-p'^{2}\right)\left( p\cdot p^{\prime}\right).
\end{align*}
Here, the global factor $\zeta$ stands for%
\[
\zeta=\frac{-e^{2}}{128\pi^{2}\left(  \left(  p\cdot p^{\prime}\right)
^{2}-p^{2}p^{\prime}{}^{2}\right)  ^{2}}.
\]


\begin{thebibliography}{99}
\bibitem{superluminal} 
%\bibitem {sudarshan1}
%\cite{Johnson:1960vt}
%\bibitem{Johnson:1960vt}
K.\ Johnson, E.\ C.\ Sudarshan,
%``Inconsistency Of The Local Field Theory Of Charged Spin 3/2 Particles,''
Annals of Physics \ {\bf 13}, 126 (1961);
%%CITATION = APNYA,13,126;%%
%\bibitem {VZ1}
G.\ Velo, D.\ Zwanziger,
%``Propagation And Quantization Of Rarita-Schwinger Waves In
%An External Electromagnetic Potential,''
Phys.\ Rev.\ {\bf 186}, 1337 (1969);
%\bibitem {VZ2}G.\ 
Velo, D.\ Zwanziger,
%" Noncausality and Other Defects of Interaction Lagrangians fo
%Particles with Spin One and Higher"
Phys. Rev. {\bf 188}, 2218 (1969).

\bibitem{NKR}
%\cite{Napsuciale:2006wr}
%\bibitem{Napsuciale:2006wr}
  M.~Napsuciale, M.~Kirchbach, S.~Rodriguez,
  %``Spin 3/2 Beyond the Rarita-Schwinger Framework,''
  Eur.\ Phys.\ J.\  {\bf A29}, 289-306 (2006).
  [hep-ph/0606308].
  
\bibitem{DN}
%\cite{DelgadoAcosta:2009ic}
%\bibitem{DelgadoAcosta:2009ic}
  E.~G.~Delgado-Acosta, M.~Napsuciale,
  %``Compton scattering off elementary spin 3/2 particles,''
  Phys.\ Rev.\  {\bf D80}, 054002 (2009).
  [arXiv:0907.1124 [hep-th]].
  
\bibitem{NRDK}
%\cite{Napsuciale:2007ry}
%\bibitem{Napsuciale:2007ry}
  M.~Napsuciale, S.~Rodriguez, E.~G.~Delgado-Acosta, M.~Kirchbach,
  %``Electromagnetic couplings of elementary vector particles,''
  Phys.\ Rev.\  {\bf D77}, 014009 (2008).
  [arXiv:0711.4162 [hep-ph]].

\bibitem{DNR}
%\cite{DelgadoAcosta:2010nx}
%\bibitem{DelgadoAcosta:2010nx}
  E.~G.~Delgado-Acosta, M.~Napsuciale, S.~Rodriguez,
  %``Second order formalism for spin 1/2 fermions and Compton scattering,''
  Phys.\ Rev.\  {\bf D83}, 073001 (2011).
  [arXiv:1012.4130 [hep-ph]].

\bibitem{feynman1}
%\cite{3828}
%\bibitem{3828} 
  R.~P.~Feynman,
  %``An Operator calculus having applications in quantum electrodynamics,''
  Phys.\ Rev.\ \ {\bf 84}, 108  (1951).
  %%CITATION = PHRVA,84,108;%%

\bibitem{Fock}
V. Fock,  Physik. Z. Sowjetunion {\bf 12}, 404 (1937). 

\bibitem{fg} 
%\cite{1814}
%\bibitem{1814} 
  R.~P.~Feynman and M.~Gell-Mann,
  %``Theory of Fermi interaction,''
  Phys.\ Rev.\ \ {\bf 109}, 193  (1958).
  %%CITATION = PHRVA,109,193;%%
  
\bibitem{schubert} 
%\cite{hep-th/0101036}
%\bibitem{hep-th/0101036} 
  C.~Schubert,
  %``Perturbative quantum field theory in the string inspired formalism,''
  Phys.\ Rept.\ \ {\bf 355}, 73  (2001)
  [hep-th/0101036].
  %%CITATION = PRPLC,355,73;%%

\bibitem{hanfeygellclasico} 
%\cite{74030}
%\bibitem{74030} 
  L.~C.~Biedenharn, M.~Y.~Han and H.~Van Dam,
  %``Two-component alternative to dirac's equation,''
  Phys.\ Rev.\ D\ {\bf 6}, 500  (1972).
  %%CITATION = PHRVA,D6,500;%%

\bibitem{cufarofeygellclasico1} 
%\cite{220844}
%\bibitem{220844} 
  N.~Cufaro Petroni, P.~Gueret, J.~P.~Vigier and A.~Kyprianidis,
  %``Second Order Wave Equation For Spin 1/2 Fields,''
  Phys.\ Rev.\ D\ {\bf 31}, 3157  (1985);
  %%CITATION = PHRVA,D31,3157;%%
%\cite{231723}
%\bibitem{231723} 
%  N.~Cufaro Petroni, P.~Gueret, J.~P.~Vigier and A.~Kyprianidis,
  %``Second Order Wave Equation For Spin 1/2 Fields. 2. The Hilbert Space Of The States,''
  Phys.\ Rev.\ D\ {\bf 33}, 1674  (1986).
  %%CITATION = PHRVA,D33,1674;%%

\bibitem{cufarofeygellclasico2}
%\cite{207362}
%\bibitem{207362} 
  N.~Cufaro Petroni, P.~Gueret and J.~P.~Vigier,
  %``Form Of A Spin Dependent Quantum Potential,''
  Phys.\ Rev.\ D\ {\bf 30}, 495  (1984).
  %%CITATION = PHRVA,D30,495;%%
 
\bibitem{hosclasico1} 
%\cite{182695}
%\bibitem{182695} 
  L.~C.~Hostler,
  %``An Sl(2,c) Invariant Representation Of The Dirac Equation,''
  J.\ Math.\ Phys.\ \ {\bf 23}, 1179  (1982).
  %%CITATION = JMAPA,23,1179;%%

\bibitem{hosclasico2}  
%\cite{195654}
%\bibitem{195654} 
  L.~C.~Hostler,
  %``An Sl(2,c) Invariant Representation Of The Dirac Equation. 2. Coulomb Green's Function,''
  J.\ Math.\ Phys.\ \ {\bf 24}, 2366  (1983).
  %%CITATION = JMAPA,24,2366;%%

\bibitem{laurie}
%\cite{47441}
%\bibitem{47441} 
  L.~M.~Brown,
  %``Two-Component Fermion Theory,''
  Phys.\ Rev.\ \ {\bf 111}, 957  (1958).
  %%CITATION = PHRVA,111,957;%%

\bibitem{tonin} 
M. Tonin, Nuovo Cimento, {\bf 14}, 1108 (1959).

  
\bibitem{hebert} 
H. Pietschmann, Acta Phys. Austr. {\bf 14}, 63 (1961).

\bibitem{3order} 
A. O. Barut and G. H. Mullen, Annals of Physics {\bf 20}, 184 (1962)

\bibitem{volkovyskii} 
%\cite{71259}
%\bibitem{71259} 
  R.~Y.~Volkovyskii,
  %``On the two-component theory of fermions,''
  Izv.\ Vuz.\ Fiz.\ \ {\bf 5}, 53  (1971);
  %%CITATION = IVUFA,5,53;%%
Russian Physics Journal {\bf 14}, 611 (1973).


\bibitem{hos1} 
%\cite{220410}
%\bibitem{220410} 
  L.~C.~Hostler,
  %``Scalar Formalism For Quantum Electrodynamics,''
  J.\ Math.\ Phys.\ \ {\bf 26}, 1348  (1985).
  %%CITATION = JMAPA,26,1348;%%
  
\bibitem{hos2} 
%\cite{237619}
%\bibitem{237619} 
  L.~C.~Hostler,
  %``Scalar Formalism For Nonabelian Gauge Theory,''
  J.\ Math.\ Phys.\ \ {\bf 27}, 2423  (1986).
  %%CITATION = JMAPA,27,2423;%%


\bibitem{Longhitano}  
%\cite{CLNS-83/562}
%\bibitem{CLNS-83/562} 
  A.~C.~Longhitano and B.~Svetitsky,
  %``Second Order Lattice Fermions,''
  Phys.\ Lett.\ B\ {\bf 126}, 259  (1983).
  %%CITATION = PHLTA,B126,259;%%  
  
\bibitem{laticepalumbohiggs4} 
%\cite{326197}
%\bibitem{326197} 
  F.~Palumbo,
  %``Second order formalism for fermions and lattice regularization,''
  Nuovo Cim.\ A\ {\bf 104}, 1851  (1991).
  %%CITATION = NUCIA,A104,1851;%%

\bibitem{morgan}  
%\cite{hep-ph/9502230}
%\bibitem{hep-ph/9502230} 
  A.~G.~Morgan,
  %``Second order fermions in gauge theories,''
  Phys.\ Lett.\ B\ {\bf 351}, 249  (1995)
  [hep-ph/9502230].
  %%CITATION = PHLTA,B351,249;%%

\bibitem{veltman} 
%\cite{hep-th/9712216}
%\bibitem{hep-th/9712216} 
  M.~J.~G.~Veltman,
  %``Two component theory and electron magnetic moment,''
  Acta Phys.\ Polon.\ B\ {\bf 29}, 783  (1998)
  [hep-th/9712216].
  %%CITATION = APPOA,B29,783;%%

\bibitem{FC} R. Mertig, http://www.feyncalc.org .


\bibitem{peskin} 
%\cite{407703}
%\bibitem{407703} 
  M.~E.~Peskin and D.~V.~Schroeder,
  ``An Introduction to quantum field theory,''
  Reading, USA: Addison-Wesley (1995).
  %%CITATION = .....,,;%%



\end{thebibliography}
\end{document}